\documentclass[%
 reprint,
 amsmath,amssymb,
 aps,
 pra,
]{revtex4-2}

\usepackage{graphicx}
\usepackage{dcolumn}
\usepackage{bm}
\usepackage{hyperref}

\newcommand{\zh}{\bm}

\newcommand{\mee}{{E}}

\newcommand{\dee}{{\varepsilon}}

\newcommand{\zhr}{{\zh r}}

\newcommand{\zhe}{{\zh e}}

\newcommand{\zhp}{{\zh p}}
\newcommand{\zhk}{{\zh k}}

\newcommand{\zhA}{{\zh A}}

\newcommand{\Br}[1]{(\ref{#1})}
\newcommand{\Eq}[1]{Eq.\ (\ref{#1})}

\newcommand{\Fig}[1]{Fig.\ \ref{#1}}
\newcommand{\txt}[1]{{\rm #1}}

\begin{document}

\title{Photon Angular Distribution in Two-Photon Electron Capture by H-Like Uranium}

\author{ Konstantin\ N.\ Lyashchenko $^{1,2,3}$}
\author{Oleg\ Yu.\ Andreev $^{1,2}$}
\author{Deyang\ Yu $^{3,4}$}

\affiliation{$^{1}$St.\ Petersburg State University, 7/9 Universitetskaya nab., St. Petersburg, 199034, Russia}
\affiliation{${}^2$Petersburg Nuclear Physics Institute named by B.P. Konstantinov of National Research Centre ``Kurchatov Institute'', mkr. Orlova roshcha 1, Gatchina, 188300, Leningrad District, Russia}
\affiliation{$^{3}$Institute of Modern Physics, Chinese Academy of Sciences, Lanzhou 730000, China}
\affiliation{$^{4}$University of Chinese Academy of Sciences, Beijing 100049, China}

\date{\today}

\begin{abstract}
We present a comprehensive study of the angular distribution of photons emitted during the resonant two-photon electron capture by H-like uranium ions. Focusing on the energies of incident electrons, at which the dielectronic recombination (DR) dominates, we analyze the angular emission spectrum of the most significant cascade transitions, which make the main contribution to the total cross section. In particular, we consider the cascade transitions that occur with the formation of $(1s2s)$ and $(1s2p)$ intermediate states. We investigate the angular distribution of the emitted photons beyond the single-photon approximation. We separately consider the contributions of the DR and the radiation recombination (RR) channels and demonstrate that the two-photon angular distribution shows strong interference between these channels.
\end{abstract}


\maketitle

\section{Introduction}
Radiative electron transitions in atoms and ions are fundamental processes in atomic physics, governing much of the behavior of matter under various physical conditions. Among these processes, one- and two-photon transitions play a important role, particularly in the emission of radiation during electron capture. Although the energy spectrum of these emissions has been widely studied \cite{Nakamura_2007,Yang2021pra,Nakamura2023PhysRevLett.130.113001,lyashchenko2024}, the angular distribution of the emitted photons, especially for two-photon processes, has received comparatively less attention. Understanding the angular distribution is crucial because it provides deeper insight into the dynamics of photon emission and the underlying atomic interactions, providing more detailed information than spectral measurements alone \cite{eichler07}.

The photon angular distribution of the one-photon radiative transitions during electron capture was extensively investigated \cite{eichler1995PhysRevA.51.3027,eichler07}. 
However, the photon angular distribution of the two-photon electron capture, in particular, involving dielectronic recombination (DR), has not been studied in detail. 
Most of the theoretical works on the DR were limited by the single-photon approximation \cite{karasiov92p453,bernhardt11,lyashchenko2015}, where only the emission of the resonant photon was considered. 
Within the framework of the single-photon approximation, it is impossible to investigate the continuous nature of the emission spectrum and take into account the complex angular correlations between the emitted photons.
In work \cite{zakowicz04}, the resonance approximation (with disregard of interference between photons) was employed for investigation of the two-photon angular distribution of the two-photon DR with H-like uranium. In particular, the angular distribution of the second photon, emitted in $(1s2p_{1/2})_1\to(1s^2)$ transition, for the DR with formation of the $(2p_{1/2}2p_{3/2})$ autoionizing states was calculated.
A comprehensive study of the full two-photon angular distribution, free from these constraints, is therefore essential for understanding the photon interaction mechanisms.

The radiative electron capture, where only the radiative recombination (RR) channel is dominant, has been thoroughly investigated in many atomic systems.
However, in systems with two or more electrons, the DR channel becomes crucial, attracting considerable attention from experimenters \cite{Nakamura_2007,Yang2021pra,Nakamura2008PhysRevLett.100.073203,bernhardt11,Mahmood_2012,Lindroth2020PhysRevA.101.062706,hu2022PhysRevA.105.L030801,Nakamura2023PhysRevLett.130.113001}. The DR is not only relevant for understanding atomic systems, but also plays a key role in describing laboratory plasmas and astrophysical phenomena \cite{hitomi2018}. The first measurements of the DR cross section were presented in \cite{Nakamura2008PhysRevLett.100.073203},\cite{bernhardt11}, where the experimental results validated theoretical predictions about the substantial influence of the Breit interaction and the interference between resonances. Recently, experimental research has been shifted to investigating the energy spectra of emitted photons in DR \cite{Nakamura_2007,Yang2021pra,Nakamura2023PhysRevLett.130.113001}. Notably, the photon emission spectra for DR in H-like and He-like iron \cite{Nakamura_2007,Yang2021pra} and Be-like lead \cite{Nakamura2023PhysRevLett.130.113001} ions were measured, focusing on the continuous nature of the two-photon emission spectrum. This was a significant step in experimental research, providing new insights into the complexities of two-photon processes and serving as motivation for this theoretical investigation.

Building on our previous work \cite{lyashchenko2024}, where we investigated the energy emission spectrum of resonant two-photon electron capture, this study focuses on the photon angular distribution of photons emitted during the same process in H-like ions. Two-photon processes exhibit rich angular correlations that reveal a complex interplay between different physical channels. Specifically, we examine the contributions of two main channels: RR and DR. The angular distribution of the emitted photons in these processes provides key insights into the nature of the intermediate states and the interactions that shape the emission pattern.

In this paper, we conduct a rigorous QED analysis of the angular distribution of photons emitted in two-photon electron capture by uranium ions. We explore how the photon angular emission patterns differ between the RR and DR channels. In particular, we demonstrate that the DR channel significantly alters the angular correlation between the emitted photons. This work represents comprehensive studies of two-photon angular distributions in highly charged ions, and it highlights the importance of considering angular correlations when studying photon emission processes.

The paper is organized as follows. In Section 2, we briefly outline the theoretical framework used to describe two-photon electron capture by H-like ions, with a more detailed description of the calculation approach available in our previous work \cite{lyashchenko2024}. Section 3 presents the energy spectrum of the emitted photons, explaining which regions of the spectrum were chosen for the analysis of the angular distribution. In Section 4, we present and analyze the calculated angular distribution of photons for resonant two-photon capture. We explore the one-photon angular distribution for each photon (with the angular variables of the other photon integrated out) and the two-photon angular distribution. The two-photon angular distribution is examined for two specific geometries: (1) where the momenta of the emitted photons are perpendicular to the momentum of the incident electron (XY-plane); and (2) where the momenta of both photons and the incident electron lie within the same plane (XZ-plane).

\section{Method of calculation} \label{theory_sec}
We perform a calculation of the differential cross section of the two-photon electron capture by a one-electron ion. 
The considered process is schematically described as
\begin{eqnarray}
e^{-}(\varepsilon)+\txt{U}^{91+}(1s)
&\to&\label{eqn170705n01x}
\cdot\cdot\cdot
\,\to\,
\txt{U}^{90+}(1s)^2 + \gamma+\gamma'
\,,
\end{eqnarray}
where the initial state is the incident electron with the energy ($\dee$) and the one-electron ion in the ground $1s$ state.
The final state is the two-electron ion in the ground $(1s)^2$ state and the two emitted photons.
Special attention is paid to the energy region of the emitted photons where the electron capture proceeds through the formation of one of the singly excited states.
In this case, the process can be described as
\begin{eqnarray}
e^{-}(\varepsilon)+\txt{U}^{91+}(1s)
&\to&\nonumber
\txt{U}^{90+}(1s,\,nl) + \gamma \\
&&\nonumber
\hspace{30pt}\downarrow
\\
&&\label{eqn170705n01}
\txt{U}^{90+}(1s)^2 + \gamma+\gamma'
\,.
\end{eqnarray}
In the DR channel, this process proceeds through the additional formation of a doubly excited state
\begin{eqnarray}
&&\nonumber
e^{-}(\varepsilon)+\txt{U}^{91+}(1s)
\,\to\,
\txt{U}^{90+}(nl,n'l')\,\to
\\
&\to&\label{eqn170705n00}
\txt{U}^{90+}(1s,n''l'') + \gamma
\,\to\,
\txt{U}^{90+}(1s)^2 + \gamma+\gamma'
\,,
\end{eqnarray}
where $n,n',n''\ge 2$. 
The two-photon electron capture process \Br{eqn170705n01x} including its subprocesses given by \Br{eqn170705n01} and \Br{eqn170705n00} is treated uniformly as a composite process.

Relativistic units are used throughout the paper unless otherwise stated.

The calculation method for two-photon electron capture within QED theory was presented in a previous work \cite{lyashchenko2024}. Here, we focus only on the key aspects of that method. The cross section for two-photon electron capture was calculated using the line profile approach (LPA), with a detailed description of LPA provided in \cite{andreev08pr}. In \cite{lyashchenko2024}, this method was further generalized to accommodate two-photon electron capture.

The two-electron wave functions of the final $(1s)^2$ state and all the intermediate states in the zero-order perturbation theory are expressed in the $j$--$j$ coupling scheme 
\begin{eqnarray}
\Psi^{(0)}_{JMn_1j_1l_1n_2j_2l_2}(\zhr_1,\zhr_2)
&=&\nonumber
N\sum\limits_{m_1m_2}
C^{JM}_{j_1m_1j_2m_2}\\
&&\label{jjcs}
\hspace{-60pt}
\times
\det\{\psi_{n_1j_1l_1m_1}(\zhr_1),\psi_{n_2j_2l_2m_2}(\zhr_2)\}
\,,
\end{eqnarray}
where the one-electron wave functions $\psi_{njlm}$ are solutions of the Dirac equation with quantum numbers $n$, $j$, $l$, and $m$, representing the principal quantum number (or the energy, for continuum electrons), total angular momentum, orbital angular momentum of the upper component, and the projection of the total angular momentum, respectively. In this expression, $J$ and $M$ represent the total angular momentum of the two-electron configuration and its projection, while 
$N$ is the normalization constant, equal to $1/\sqrt{2}$ for nonequivalent electrons and $1/2$ for equivalent electrons. The symbols $C^{JM}_{j_1m_1j_2m_2}$ are the Clebsch-Gordan coefficients \cite{Varshalovich1988QuantumTO}.

The initial state of the electron system ($1s$ and $e^{-}(\varepsilon)$) includes the incident electron with a certain momentum ${\zhp}$ and polarization $\mu$ (in the asymptotic $r\to\infty$). Its wave function can be written as
\begin{eqnarray}
\Psi^{(0)}_{njlm, {\zhp} \mu}(\zhr_1,\zhr_2)
&=&
\frac{1}{\sqrt{2}} \det\{ \psi_{njlm}(\zhr_1), \psi_{{\zhp} \mu}(\zhr_2) \}
\,,
\end{eqnarray}
where $\psi_{njlm}$ is the wave function of the bound electron, $\psi_{{\zhp} \mu}$ is the wave function of the continuum electron.

The interaction between electrons plays a crucial role in the formation of doubly excited (autoionizing) states and must therefore be properly accounted for. These doubly excited states are typically quasidegenerate, so quasidegenerate QED perturbation theory should be applied to accurately describe the DR channel.

For this purpose, we use the LPA. Within the framework of the LPA, the function describing a two-electron state, which accounts for the interaction of quantized electromagnetic and electron-positron fields, takes the following form \cite{lyashchenko2024, andreev08pr, andreev09p042514}
\begin{eqnarray}
\Phi_{K}
&=&\label{eigenphi}
\sum\limits_{ N}
B_{N {K}}\Psi^{(0)}_{ N}
\end{eqnarray}
where $K\equiv(JMj_1j_2l_1l_2n_1n_2)$ is a composite index representing the complete set of quantum numbers describing the reference state $K$, and the index $N$ runs over all two-electron configurations, including integration over both positive- and negative-energy continua. In this work, the functions $\Phi_{K}$ account for the electron self-energy corrections, vacuum polarization corrections, and one- and two-photon exchange corrections.

Within the LPA the amplitude of the two-photon electron capture can be expressed as \cite{andreev08pr}
\begin{eqnarray}
U_{FI}
&=&\nonumber
\sum_{N}\frac{\left(A^{(k_1,\lambda_1)*}\right)_{FN} \left(A^{(k_2,\lambda_2)*}\right)_{NI} }{E_{F}+\omega_1-E_{N}+\frac{i}{2}\Gamma_{N}}
\\
&+&\label{3}
\sum_{N}\frac{\left(A^{(k_2,\lambda_2)*}\right)_{FN} \left(A^{(k_1,\lambda_1)*}\right)_{NI} }{E_{F}+\omega_2-E_{N}+\frac{i}{2}\Gamma_{N}}
\,,
\end{eqnarray}
where indices $I$, $N$ and $F$ denote the initial, intermediate and final two-electron states with energies $\mee_I$, $\mee_N$ and $\mee_F$, respectively. The width of an intermediate state is represented by $\Gamma_{N}$. The two-electron matrix element of the photon emission $\left(A^{(k,\lambda)*}\right)_{UD}$ reads as
\begin{eqnarray}
\left(A^{(k,\lambda)*}\right)_{UD}
&=&\nonumber
e\int d^3 {\bf r}_1 d^3 {\bf r}_2 \,
\overline{\Phi}_{U} ({\bf r}_1, {\bf r}_2)  
\\
&&\nonumber
\hspace{-40pt}
\times
\left(
\gamma^{(1)\nu}A^{(k,\lambda)*}_{\nu}(\zhr_1)
+
\gamma^{(2)\nu}A^{(k,\lambda)*}_{\nu}(\zhr_2)
\right)
\\
&&\label{Eq576789}
\hspace{-40pt}
\times
\Phi_{D} ({\bf r}_1, {\bf r}_2)
\,.
\end{eqnarray}
In \Eq{Eq576789} $\gamma^{(i)\nu}$ are the Dirac $\gamma$ matrices that act on the one-electron wave function of the argument $\zhr_i$.
The photon wave function
$A^{(k,\lambda)\nu}=(A_0^{(k,\lambda)},{\zhA}^{(k,\lambda)})$
in the transverse gauge reads as
\begin{eqnarray}
A_0^{(k,\lambda)}(\zhr)
&=&
0
\,,\,\,\,
{\zhA}^{(k,\lambda)}({\zhr})
\,=\,
\sqrt{\frac{2\pi}{\omega}} e^{i \zhk\zhr} \zhe^{(\lambda)}
\,,
\end{eqnarray}
where $\zhk$ is the photon wave vector, $\omega=|\zhk|$ is the photon energy (frequency),
$\zhe^{(\lambda)}$ is the polarization vector.

The summations in \Eq{3} run over the complete basis set constructed from the two-electron functions \Eq{eigenphi}.
However, in this work, it is sufficient to take into account only the two-electron $(n_1 l_1, \, n_2 l_2)$ states where the principal quantum number $n_1$ of the first electron is equal $1$ and $2$.
The quantum numbers of the second electron run over the complete Dirac spectrum.

The fully differential cross section is connected with the amplitude as
\begin{eqnarray}
\frac{d^4\sigma}{d\omega_1 d\Omega_1 d\omega_2 d\Omega_2}
&=&\nonumber
\delta(\omega_1+\omega_2-E_{I}+E_{F})\\
&\times&\label{2}
\left|U_{FI}\right|^2
\frac{\varepsilon}{p}
\frac{\omega_1^2 \omega_2^2}{(2\pi)^5}
\,,
\end{eqnarray}
where $\Omega_{1,2}$ are the solid angles of the emitted photons.

In this work we consider unpolarized electrons and photons which means that averaging over electron polarizations $\mu$ and $m_b$ as well as summation over photon polarization $\lambda_1$ and $\lambda_2$ should be performed:  
\begin{eqnarray}
\frac{d^3\sigma}{d\omega_1 d\Omega_1 d\Omega_2}
&=&
\frac{1}{4}\sum_{\mu,m_b}
\!
\sum_{\lambda_1,\lambda_2} 
\!
\int d\omega_2 
\!
\frac{d^4\sigma}{d\omega_1 d\Omega_1 d\omega_2 d\Omega_2} 
\,.
\end{eqnarray}

\section{Energy distribution}
We investigate the process of electron capture by a hydrogen-like uranium ion initially in the 1s state, accompanied by two-photon emission. The incident electron energy is chosen so that the energy of the initial state ($E_{I}=\epsilon(e)+\epsilon(1s)$) is close to the energy of one of the doubly excited states (such as $(2s2s)$, $(2s2p)$, or $(2p2p)$). In this scenario, dielectronic recombination plays a significant role. Specifically, electron capture can occur through the formation of doubly excited states, followed by their subsequent decay.

The singly differential cross section of the electron capture with emission of the two photons is given by
\begin{eqnarray}
\frac{d\sigma}{d\omega_1}
&=&\label{eq_983}
\int d\Omega_1 d\Omega_2
\frac{d^3\sigma}{d\omega_1 d\Omega_1 d\Omega_2}
\,.
\end{eqnarray}
The energies of the emitted photons are lie in the interval determined by the energy conservation law
\begin{eqnarray}
\omega_1+\omega_2
&=&\label{eqn231124n01}
E_{I}-E_{F}
\,.
\end{eqnarray}
Accordingly, the photon energy spectrum is continuous and limited by the energy interval $[0,\omega_{\txt{max}}]$, where
$\omega_{\txt{max}}=E_{I}-E_{F}$.
We note that if one of the photons is registered, then the energy of the other is determined by
\Eq{eqn231124n01}.

\begin{figure}[ht]
\includegraphics[width=20pc]{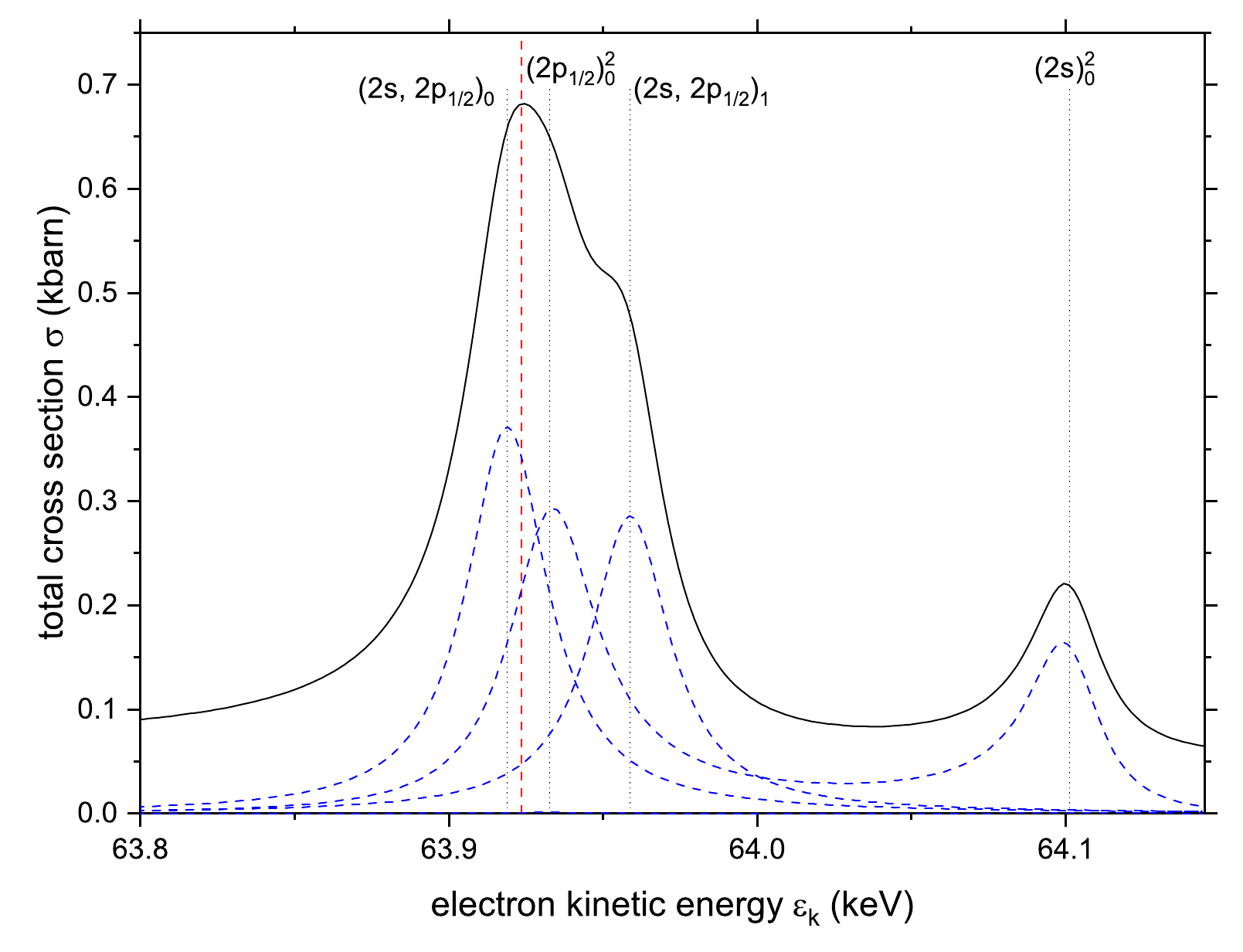}
\caption{Total cross section of the two-photon electron capture by H-like uranium ion as a function of incident electron kinetic energy.}
\label{fig_1}
\end{figure}

The total cross section reads as 
\begin{eqnarray}
\sigma
&=&
\int^{\frac{\omega_{\txt{max}}}{2}}_{0} d\omega_1
\frac{d\sigma}{d\omega_1}
\,.
\end{eqnarray}
In Fig. \ref{fig_1} we present the total cross section of two-photon electron capture as a function of incident electron kinetic energy. The resonances in the cross section indicate a strong contribution from dielectronic recombination, manifesting in the formation of the $(2s2p_{1/2})_0$, $(2p_{1/2})^2$, $(2s2p_{1/2})_1$, and $(2s)^2$ autoionizing states. 
The contributions from these autoionizing states are represented by a blue dotted line.
A detailed study of the total cross section of this process can be found in \cite{karasiov92p453,zakowicz04,andreev09p042514,lyashchenko2015,lyashchenko2024}.
For further investigation of the differential cross section we choose the electron kinetic energy equal to 63.9235 keV, corresponding to the maximum of the total cross section. This energy is marked by the red dashed vertical line in  Fig. \ref{fig_1}. For this collision energy, Eq.\~(\ref{eqn231124n01}) has the form
\begin{eqnarray}
\omega_1+\omega_2
&=&
193.4885 \,\txt{keV}
\,.
\end{eqnarray}

\begin{figure}[ht]
\includegraphics[width=20pc]{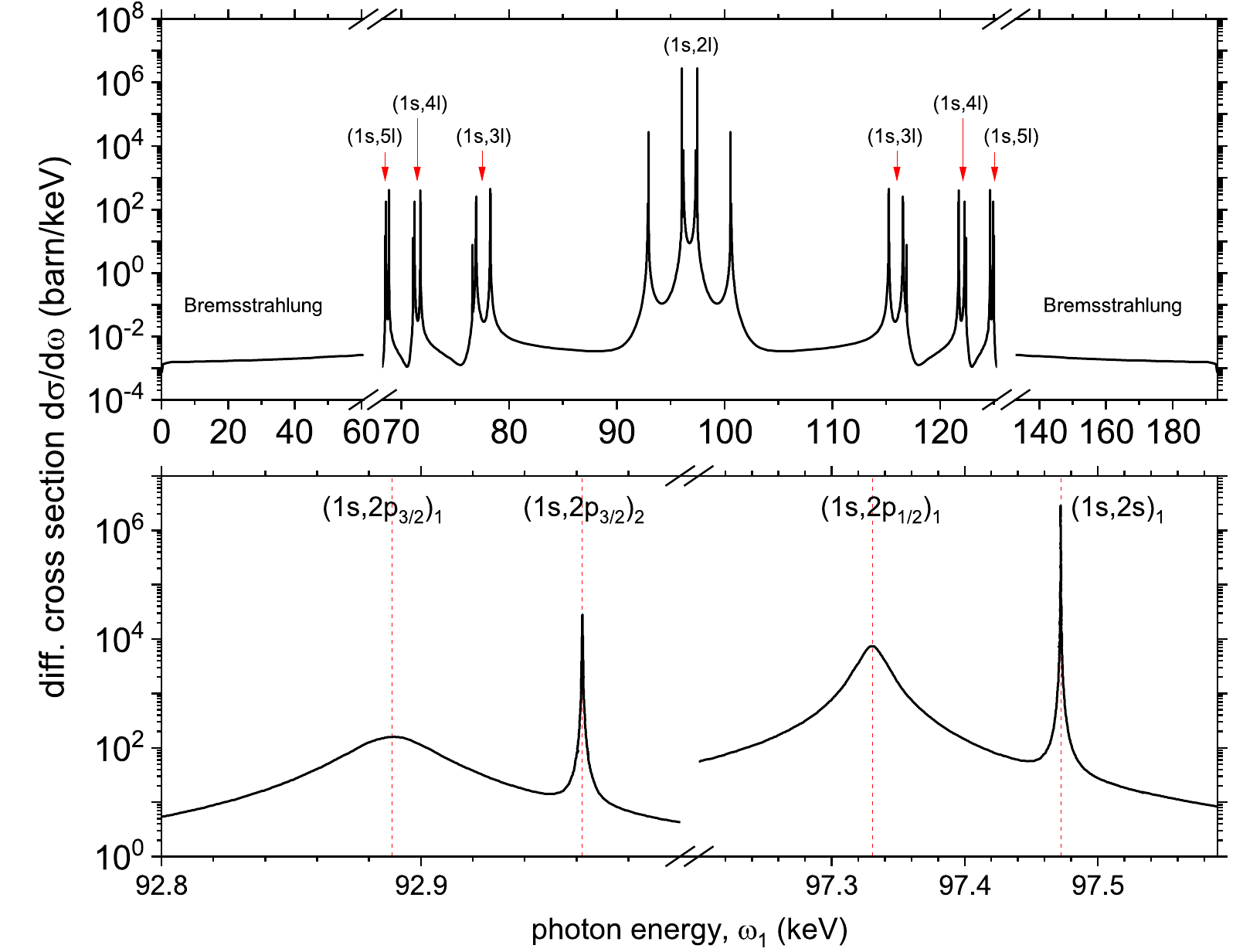}
\caption{Differential cross section of the two-photon electron capture by H-like uranium ion as a function of the photon energy ($\omega_1$) of one of the emitted photons (other photon energy $\omega_2=\omega_{\txt{max}} - \omega_1$). The kinetic energy of the incident electron equal to 63.9235 keV. The red dotted lines in the lower panel mark the positions of the cascade resonances with $(1s2l)$ states (see Table \ref{table0}).}
\label{fig_2}
\end{figure}

In Fig. \ref{fig_2}, we present a differential cross section of two-photon electron capture as a function of energy of one of the emitted photons for chosen incident electron kinetic energy. 
Detailed investigation of the photon energy distribution was performed in our previous work \cite{lyashchenko2024}.
Since the two-photon states $(\zhk_1,\lambda_1;\zhk_2,\lambda_2)$ and $(\zhk_2,\lambda_2;\zhk_1,\lambda_1)$ are identical, the differential cross section \Eq{eq_983} is symmetric with respect to the center of the energy interval $[0,\omega_{\txt{max}}]$.
The two-photon electron capture predominantly proceeds via cascade transitions $(1s,e)\to(1s,nl)\to(1s)^2$ with the formation of singly excited $(1s,nl)$ states with $n\ge2$. 
In particular, the main contributions come from the energy regions corresponding to the cascade transitions through the  $(1s2s)$ and $(1s2p)$ states. Therefore, we focus on the four pairs of photon energies ($\omega_1^{{\txt{res}},N}$, $\omega_2^{{\txt{res}},N}$), which are associated with the formation and radiative decay of four singly excited states $N$: $(1s2s)_{1}$, $(1s2p_{1/2})_{1}$, $(1s2p_{3/2})_{1}$, and $(1s2p_{3/2})_{2}$. These energies are presented in Table \ref{table0} and shown by the vertical red dotted lines in the lower panel of Fig.~\ref{fig_2}.

Strictly speaking, in a two-photon process, we cannot assign a specific photon to the particular transition. However, the two photons can be distinguished by their energies. In the lower panel of Fig.~\ref{fig_2}, we consider photon energy regions that include the energy differences between the initial state and singly excited states ($E_I-E_{(1s2l)}$). The energy of the other photon (determined by the energy conservation law) is close to the energy difference between a singly excited state and the final state ($E_{(1s2l)}-E_F$). These two photons are therefore distinguishable by their energies. We refer to the first photon $(\omega_1,\zhk_1)$ as the resonant photon with the energy corresponding to the $(1s,e)\to(1s,2l)$ transition, and the second photon $(\omega_2,\zhk_2)$ as the satellite photon with the energy corresponding to the $(1s,2l)\to(1s,1s)$ transition.

\begin{table}
\caption{The cascade energies of the emitted photons $\omega_1^{{\txt{res}},N}=E_{I}-E_{N}$ and $\omega_2^{{\txt{res}},N}=E_{N}-E_{F}$ as well as the natural line widths $\Gamma_{N}$ of the corresponding single excited states.}
\begin{ruledtabular}
\begin{tabular}{ c| c| c| c}
 state & $\omega_1^{{\txt{res}},N}$ & $\omega_2^{{\txt{res}},N}$ & $\Gamma_{N}$ \\ 
 N & (keV) & (keV) & (eV)\\ 
  \hline
 $(1s2s)_1$ & 97.4668 & 96.0217 & 0.09\\  
 $(1s2p_{1/2})_1$ & 97.3257 & 96.1628 & 20.3\\ 
 $(1s2p_{3/2})_1$ & 92.8835 & 100.605 & 34.0\\  
 $(1s2p_{3/2})_2$ & 92.9572 & 100.5313 & 0.22
\end{tabular}
\end{ruledtabular}
\label{table0}
\end{table}

\section{Angular distribution}
We align the $z$-axis with the momentum of the incident electron, and measure the polar angle $\theta$ from this axis. The photon momentum is described by the photon energy $\omega$, the polar angle $\theta$ and the azimuthal angle $\varphi$: $\zhk=\zhk(\omega,\theta,\varphi)$.
The differential cross section for two-photon ($\zhk_1$ and $\zhk_2$) emission is characterized by the angles of both photons. However, because of the presence of axial symmetry, the differential cross section depends only on the difference in the azimuthal angles
\begin{eqnarray}
    \frac{d^3\sigma}{d\omega_1 d\Omega_1 d\Omega_2}
    &=&\label{Eq_diff}
    \frac{d^3\sigma}{d\omega_1 d\Omega_1 d\Omega_2}(\omega_1,\theta_1,\theta_2,|\varphi_2-\varphi_1|)
    \,.
\end{eqnarray}

It is worth noting that, due to the identity of the two pairs ($\zhk_1$, $\zhk_2$) and ($\zhk_2$, $\zhk_1$) of emitted photons, the following equality holds:
\begin{eqnarray}
    \frac{d^3\sigma}{d\omega_1 d\Omega_1 d\Omega_2}(\omega_1,\theta_1,\theta_2,|\varphi_2-\varphi_1|)
    &&\nonumber
    \\
    &&\hspace{-4.5cm}
    =\frac{d^3\sigma}{d\omega_1 d\Omega_1 d\Omega_2}(\omega_{\txt{max}}-\omega_1,\theta_2,\theta_1,|\varphi_2-\varphi_1|)
    \,.
\end{eqnarray}

\subsection{One-photon angular distribution} \label{subsec_one_ph_ang_distr}
\begin{figure}[ht]
\includegraphics[width=20pc]{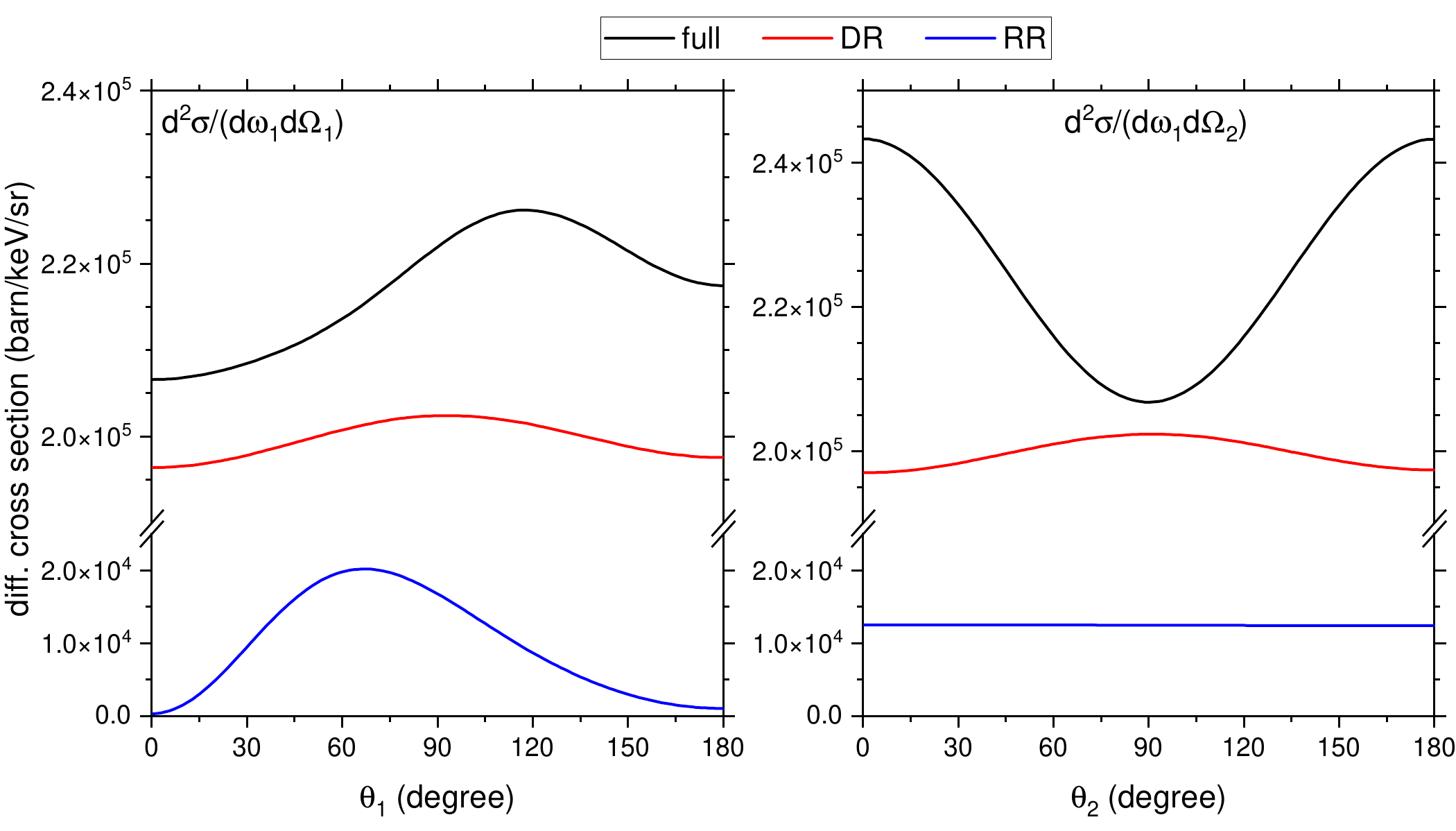}
\caption{Differential cross section as a function of polar angle of one of the emitted photons ($\theta_1$ or $\theta_2$). The photon energy corresponds to $(1s2s)_1$ cascade resonance.
The left panels correspond to the first photon with energy $\omega_1 = E_I - E_{(1s2s)_1}=97.4668$ keV, and the right panels correspond to the second photon with energy $\omega_2 = E_{(1s2s)_1} - E_F=96.0217$ keV.
The black curve shows the result of the full calculation, the red curve corresponds to the DR contribution, and the blue curve demonstrates the RR contribution.}
\label{fig_one_ph_5}
\end{figure}

\begin{figure}[ht]
\includegraphics[width=20pc]{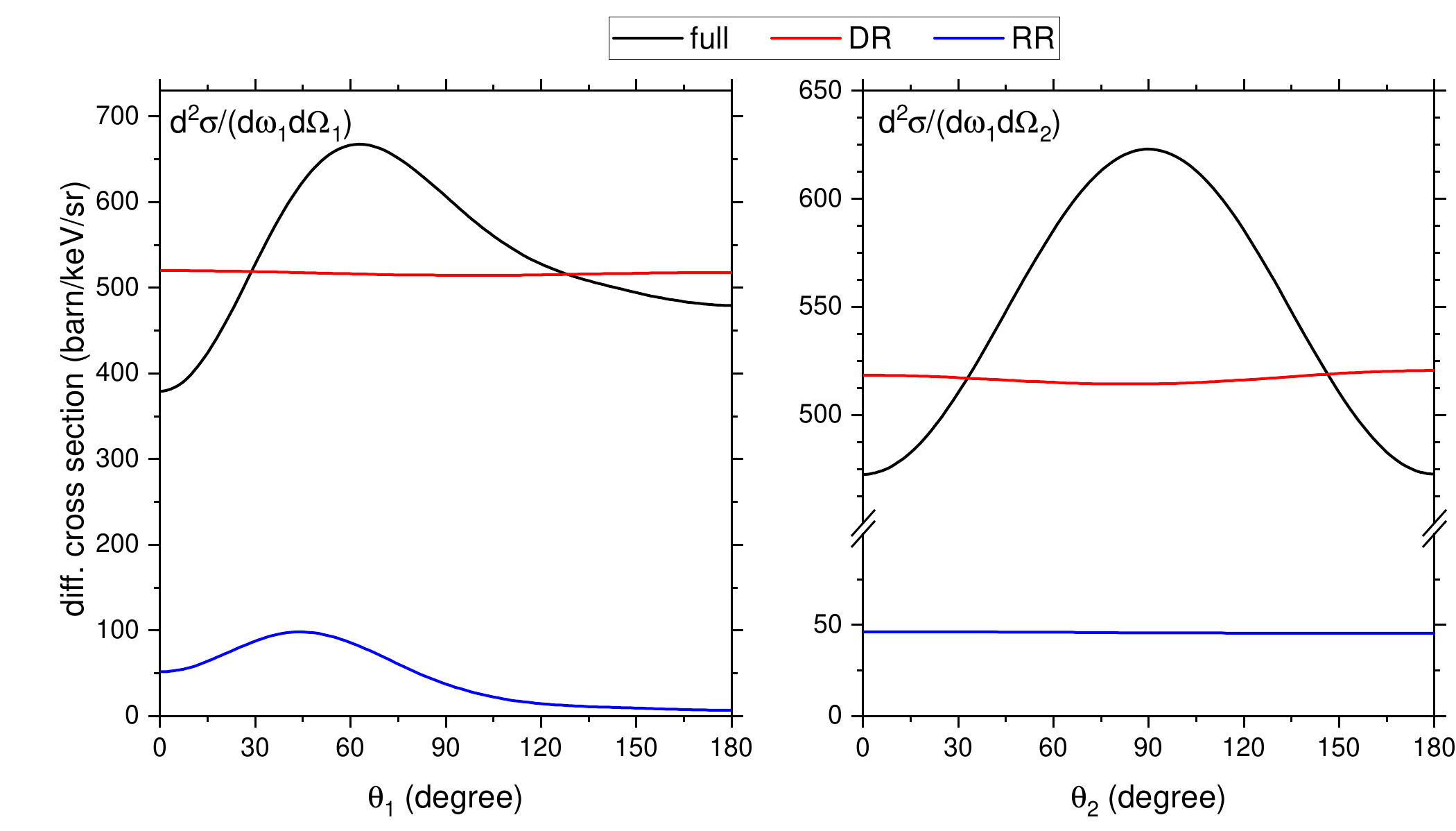}
\caption{The same as in Fig. \ref{fig_one_ph_5}, but for photon energy corresponding to $(1s2p_{1/2})_1$ cascade resonance: $\omega_1=97.3257$ keV, $\omega_2=96.1628$ keV.}
\label{fig_one_ph_6}
\end{figure}

\begin{figure}[ht]
\includegraphics[width=20pc]{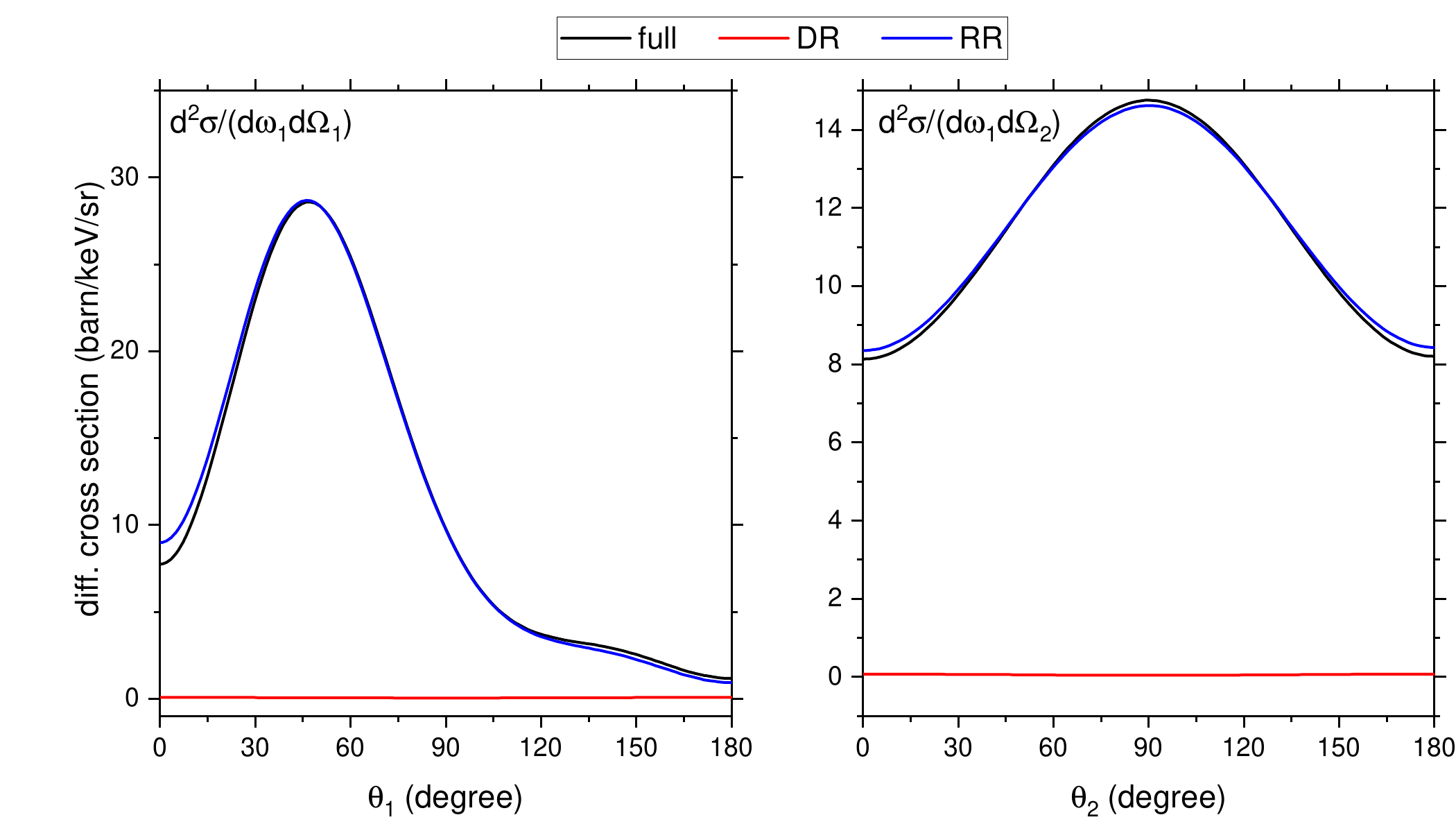}
\caption{The same as in Fig. \ref{fig_one_ph_5}, but for photon energy corresponding to $(1s2p_{3/2})_1$ cascade resonance: $\omega_1=92.8835$ keV, $\omega_2=100.605$ keV.}
\label{fig_one_ph_7}
\end{figure}

\begin{figure}[ht]
\includegraphics[width=20pc]{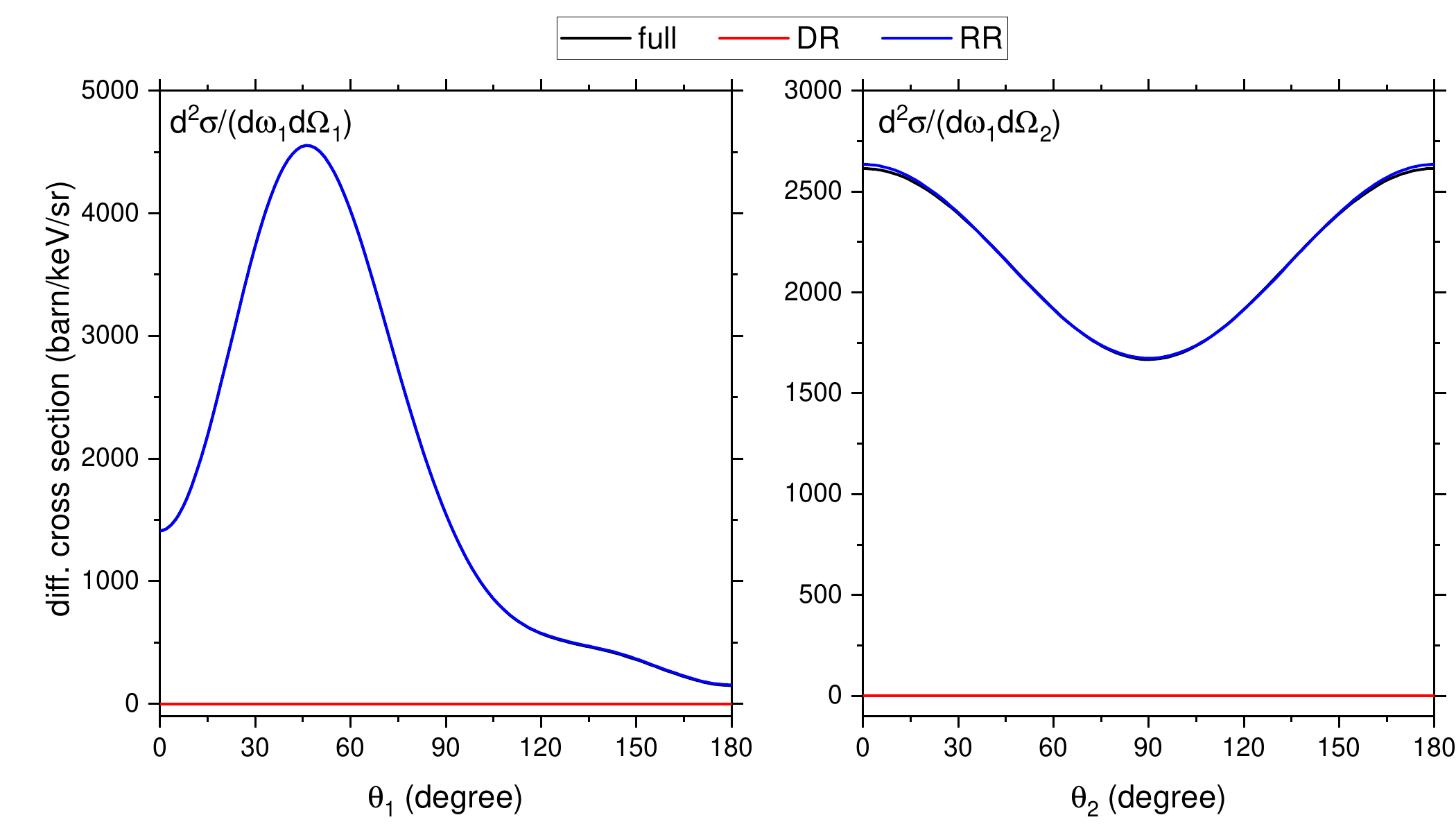}
\caption{The same as in Fig. \ref{fig_one_ph_5}, but for photon energy corresponding to $(1s2p_{3/2})_2$ cascade resonance: $\omega_1=92.9572$ keV, $\omega_2=100.5313$ keV.}
\label{fig_one_ph_8}
\end{figure}

Here, we assume that only one (either $(\omega_1,\zhk_1)$ or $(\omega_2,\zhk_2)$) of the emitted photons is detected, while the other photon is not.
Accordingly, for investigation of the one-photon angular distribution, we consider the differential cross section in which integration over the angular variables of the first or second emitted photons is performed, respectively,
\begin{eqnarray}
\frac{d^2\sigma}{d\omega_1 d\Omega_1}
&=&\label{Eq_735}
\int d\Omega_2
\frac{d^3\sigma}{d\omega_1 d\Omega_1 d\Omega_2} 
\,,
\\
\frac{d^2\sigma}{d\omega_1 d\Omega_2}
&=&\label{Eq_737}
\int d\Omega_1
\frac{d^3\sigma}{d\omega_1 d\Omega_1 d\Omega_2} 
\,.
\end{eqnarray}
We note that these differential cross sections are independent of the azimuthal photon angles ($\varphi_i$).
The differential cross sections as functions of the polar angle ($\theta_i$) of the first or second emitted photons are presented in Figs. \ref{fig_one_ph_5} - \ref{fig_one_ph_8}, where the photon energies are $\omega_1=E_I-E_{N}$ and $\omega_2=E_{N}-E_F$, with $N$ representing one of the $(1s 2s)_1$, $(1s 2p_{1/2})_1$, $(1s 2p_{3/2})_1$, and $(1s 2p_{3/2})_2$ states, respectively. The contributions from the DR and RR channels are shown separately. The DR channel is dominant for the $(1s2s)_1$ and $(1s2p_{1/2})_1$ cascades (Figs. \ref{fig_one_ph_5} and \ref{fig_one_ph_6}), while the RR channel is dominant for the $(1s2p_{3/2})_{1,2}$ cascades (Figs. \ref{fig_one_ph_7} and \ref{fig_one_ph_8}). We observe that the significant contribution from the DR channel results in the differential cross section for the DR resonances being substantially larger than that for the RR resonances. 
These figures indicate that the shape of the angular distribution depends on the relative contributions of the RR and DR channels. In the RR channel, the first photon is typically emitted in the forward direction, regardless of the resonance. In contrast, in the DR channel, the angular distribution of the first emitted photon can vary significantly depending on the specific autoionizing state formed. The second photon, in both channels, has a symmetrical distribution with respect to $90^\circ$.

These figures also show evidence of strong interference between the RR and DR channels. The cascade contribution of the second photon is associated with the $(1s2l)\to(1s1s)$ transitions in both channels. The corresponding RR and DR amplitudes differ only by a phase shift, leading to significant interference in transitions where both the RR and DR channels are strong. For the first photon, the situation is different: in the RR channel, the cascade contribution comes from the $(1s,e)\to(1s,2l)$ transition, while in the DR channel, it originates from the $(2s2p)\to(1s2l)$ transition. Since the corresponding RR and DR amplitudes are substantially different, the interference between the RR and DR channels for the first photon is less pronounced.

\begin{figure}[ht]
\includegraphics[width=20pc]{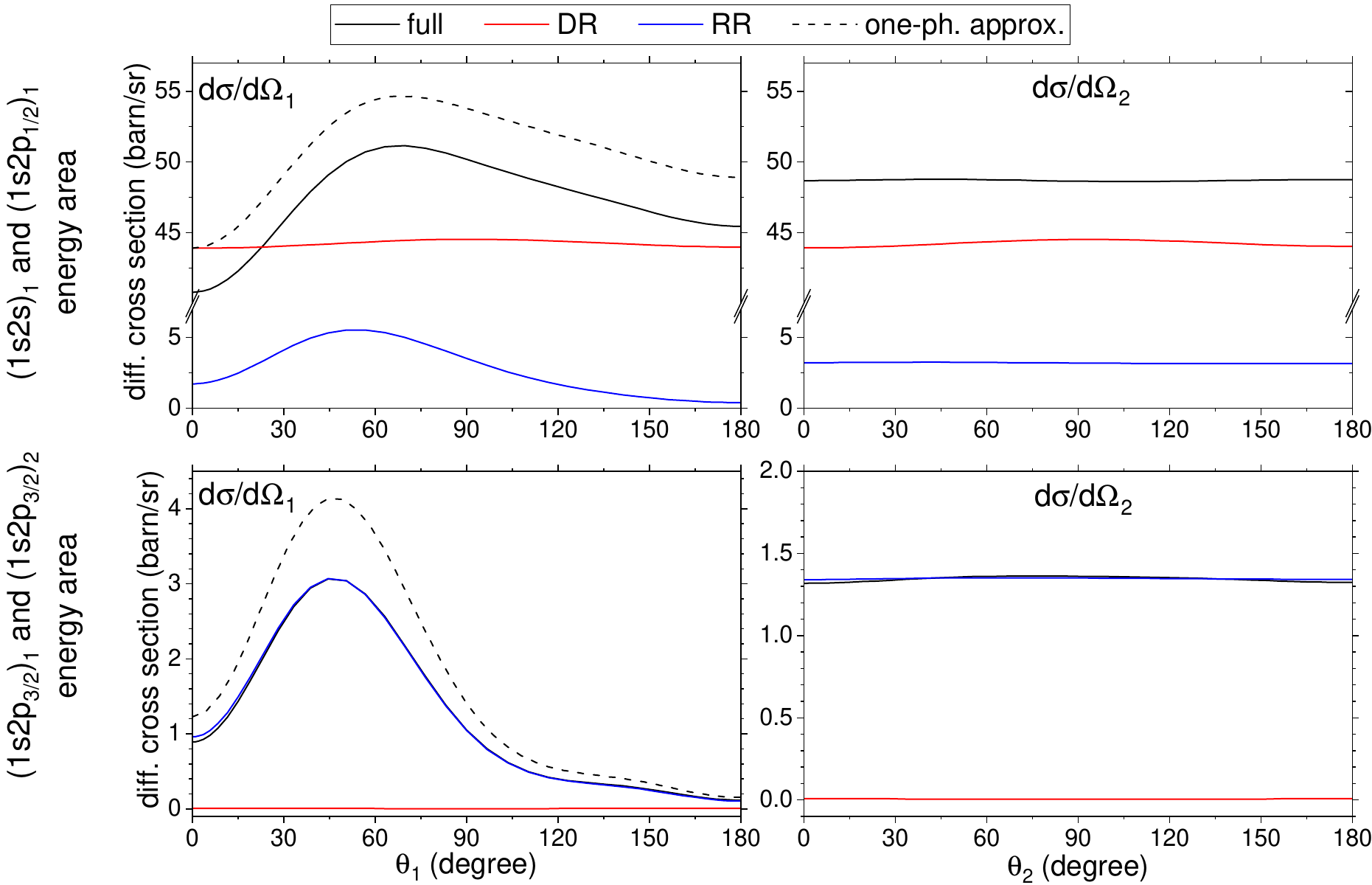}
\caption{
The differential cross section is presented as a function of the polar angle of one of the emitted photons. The left panels show the angular distribution of the first photon (with energy $\omega_1 \approx E_I - E_{(1s2l)}$), while the right panels show the angular distribution of the second photon (with energy $\omega_2 \approx E_{(1s2l)} - E_F$). The top panels display differential cross section after integration over the photon energy $\omega_1$ within [96.312, 98.339] keV interval which includes two close cascade resonances with $(1s2s)_1$ and $(1s2p_{1/2})_1$ states. The bottom panels show differential cross section after integration within [91.181, 94.585] keV interval which includes other two close cascade resonances with $(1s2p_{3/2})_1$ and $(1s2p_{3/2})_2$ states.
The black curve shows the result of the full calculation, the red curve corresponds to the DR contribution, and the blue curve demonstrates the RR contribution.  
}
\label{fig_one_ph_integ_DR_RR}
\end{figure}

To account for the accuracy of the photon detectors and the energy spread of the incident electrons in the beam, we perform integration over the energy of the emitted photons. The four considered cascade resonances can be divided into two groups of resonances that are close in energy. The first group includes cascade transitions with the formation of the $(1s2s)_1$ and $(1s 2p_{1/2})_1$ states, while the second group includes cascades transitions with formation of the $(1s 2p_{3/2})_1$ and $(1s 2p_{3/2})_2$ states. The integration energy interval is [$\omega_1^{(\txt{res},N)} - 50 \Gamma_N$, $\omega_1^{(\txt{res},N)} + 50\Gamma_N$], where $N = (1s2p_{1/2})_1$ for the first group and $N = (1s2p_{3/2})_1$ for the second group. The energies $\omega_1^{(\txt{res},N)}$ and widths $\Gamma_N$ are presented in Table~\ref{table0}.
The results of these integrations are given in Fig. \ref{fig_one_ph_integ_DR_RR}.
This figure shows that after integration over the energy intervals, the first photon still has a noticeable maximum in the angular distribution and tends to be emitted in the forward direction, while the second photon exhibits almost isotropic behavior for the both groups of resonances.
The reason for the isotropic behavior is the significant compensation of the maximum of the angular distribution for one resonance and the minimum for the other in each group. We note that in each resonance group, one of the resonances has a small maximum in energy differentiated cross section but has a large width, and the other, on the contrary, has a large energy differentiated cross section and a small width, which provides compensation leading to the isotropic distribution.

The angular distribution of the DR process within the single-photon approximation (where the angular distribution of only the first photon is available) was studied in \cite{karasiov92p453,bernhardt11,lyashchenko2015}. The angular distribution of the first photon, obtained within this approximation, is shown as black dotted curves in Fig. \ref{fig_one_ph_integ_DR_RR}. 
The angular distribution of the first photon in the full calculation and approximation is qualitatively similar, but there is a noticeable quantitative difference. We explain this difference by the following. First, the destructive interference between cascade resonances decreases the cross section. 
Second, the results of the single-photon approximation are close to the results of the exact calculation if the ratio of the partial single-photon width corresponding to the decay to the ground state ($\Gamma_{N}^{(1\gamma,N\to 1s^2)}$) to the total width ($\Gamma_{N}$) is close to unit ($\Gamma_{N}^{(1\gamma,N\to 1s^2)} / \Gamma_{N} \approx 1$) for each intermediate cascade state $N$ (see Eq. (A8) in \cite{lyashchenko2024}). This ratio is indeed close to unit for the states $(1s2s)_1$, $(1s2p_{1/2})_1$, and $(1s2p_{3/2})_1$, but for the state $(1s2p_{3/2})_2$ this ratio is 0.62 due to the presence of the single-photon $(1s2p_{3/2})_2 \to (1s2s)_1$ transition. 
Finally, the third reason is the limited interval of integration.

\subsection{Two-photon angular distribution}
Here, we assume that both of the emitted photons are detected. The process is characterized by three vectors: $\zhp=p\zhe_z$, $\zhk_1(\theta_1,\varphi_1)$, and $\zhk_2(\theta_2,\varphi_2)$, so we align the $z$-axis with the momentum of the incident electron.
To investigate the angular distribution of the emitted photons, we consider two characteristic scenarios. In the first scenario, the emitted photons momenta lie in the XY-plane. In the second scenario, the momenta lie in the XZ-plane.

\subsubsection{XY-plane}
In the scenario where both emitted photon momenta lie in the XY-plane ($\theta_1=\theta_2=90^\circ$), the differential cross section depends only on the angle between the photon momenta. 
The differential cross section $\frac{d^3\sigma}{d\omega_1 d\Omega_1 d\Omega_2}$ can be considered as a function of $\omega_1$ and $\varphi=|\varphi_2-\varphi_1|$
\begin{eqnarray}
    \frac{d^3\sigma^{XY}}{d\omega_1 d\Omega_1 d\Omega_2}\left(\omega_1,\varphi \right)  
    &&\nonumber
    \\
    &&\label{sks9309d}
    \hspace{-1.4cm} 
    =\frac{d^3\sigma}{d\omega_1 d\Omega_1 d\Omega_2}\left(\omega_1,\frac{\pi}{2},\frac{\pi}{2},|\varphi_2-\varphi_1|\right)
    \,.
\end{eqnarray}

Due to the identity of the emitted photons, the cross section has the following energy symmetry in the XY plane
\begin{eqnarray}
    \frac{d^3\sigma^{XY}}{d\omega_1 d\Omega_1 d\Omega_2}\left(\omega_1,\varphi \right) 
    &=&
    \frac{d^3\sigma^{XY}}{d\omega_1 d\Omega_1 d\Omega_2}\left(\omega_{\txt{max}}-\omega_1,\varphi \right)  
    \,.
\end{eqnarray}

\begin{figure}[ht]
\includegraphics[width=20pc]{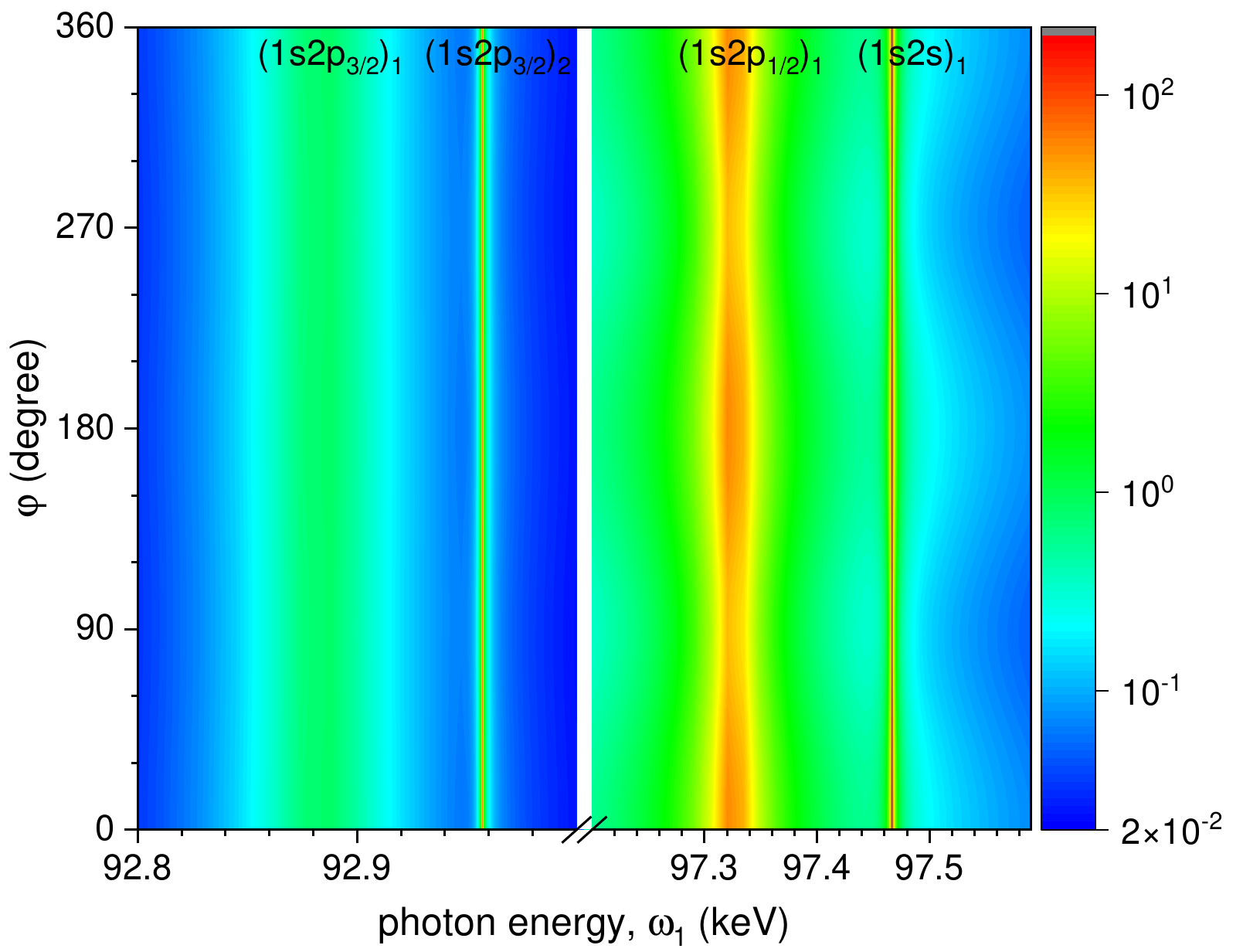}
\caption{(XY-plane)
Differential cross section (in barn/keV/sr$^2$) for two-photon electron capture by a H-like uranium ion, 
shown as a function of the energy of the first photon $\omega_1$ and the angle between photons momenta $\varphi$.
The narrow purple region, located at the $(1s2s)_1$ resonance, represents a differential cross section exceeding $200$ barn/keV/sr$^2$, with a maximum value of  $2.4 \times 10^4$ barn/keV/sr$^2$.
}
\label{fig_3}
\end{figure}

The differential cross section is shown in \Fig{fig_3}. The horizontal axis represents the energy of the first photon and the vertical axis represents the angle between photon momenta $\varphi$.
We can observe the contribution of the four singly excited states ($(1s2s)_{1}$, $(1s2p_{1/2})_{1}$, $(1s2p_{3/2})_{1}$, and $(1s2p_{3/2})_{2}$) to the angular distribution. As before, we focus on the differential cross section at fixed photon energies corresponding to the maxima of the cascade resonances marked by red lines in the lower panel of Fig. \ref{fig_2} and presented in Table \ref{table0}.

For investigation of the $\varphi$ dependence in \Eq{Eq_diff}, it is convenient to introduce the normalized differential cross section 
\begin{eqnarray}
    \frac{d^3 \tilde{\sigma}^{XY}}{d\omega_1 d\Omega_1 d\Omega_2}
    &=&\label{Eq_diff_norm}
    \frac{d^3\sigma^{XY}}{d\omega_1 d\Omega_1 d\Omega_2}
    \left/
    \frac{d^3\sigma^{XY}}{d\omega_1 d\cos\theta_1 d\cos\theta_2}
    \right.
    \,,
\end{eqnarray}
where
\begin{eqnarray}
    \frac{d^3\sigma^{XY}}{d\omega_1 d\cos\theta_1 d\cos\theta_2}
    &=&
    \int \frac{d^3\sigma^{XY}}{d\omega_1 d\Omega_1 d\Omega_2} d\varphi_1 d\varphi_2
    \,.
\end{eqnarray}

\begin{figure}[ht]
\includegraphics[width=20pc]{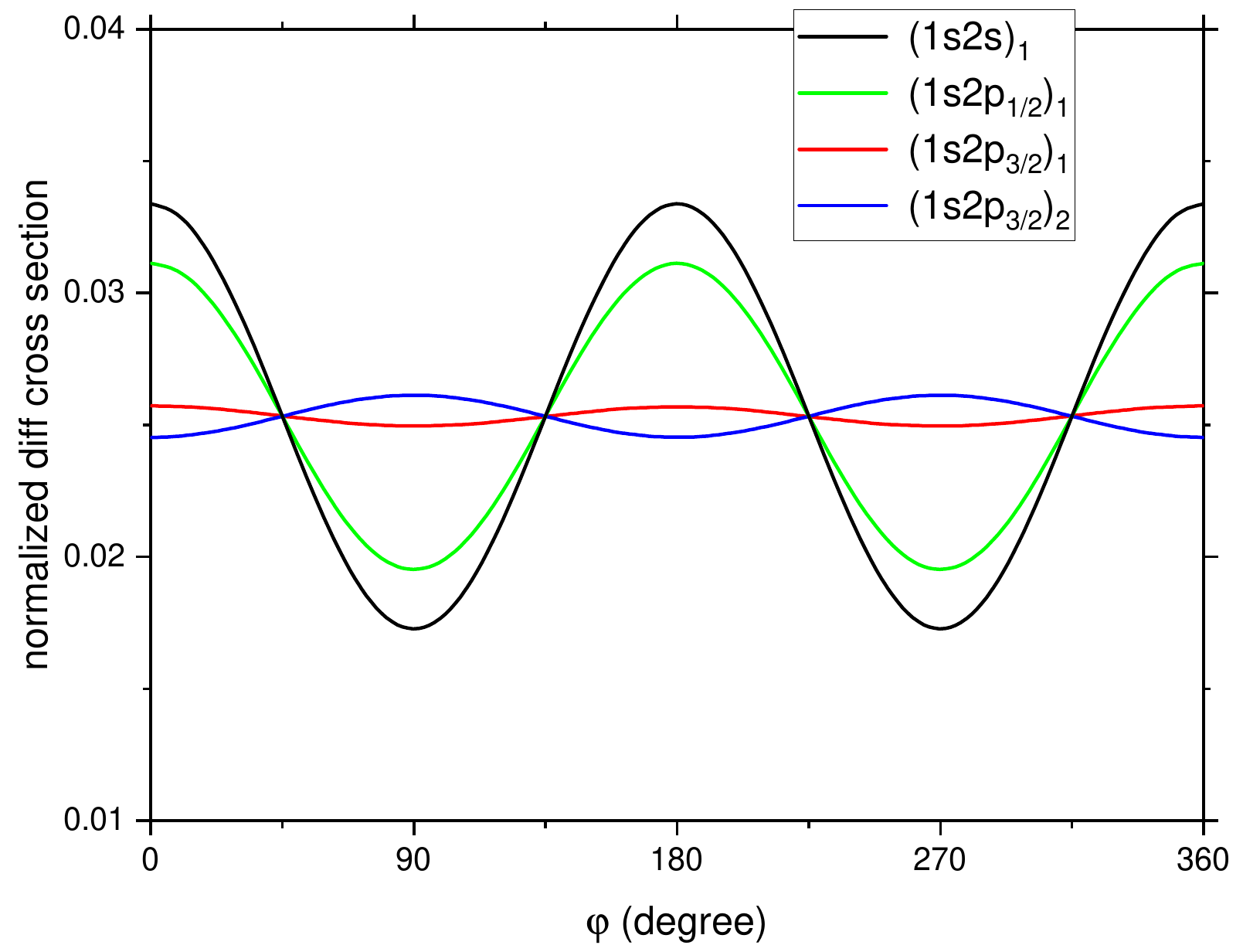}
\caption{(XY-plane) Normalized differential cross section of the two-photon electron capture by H-like uranium ion as 
a function of $\varphi=|\varphi_2-\varphi_1|$ for photon energies corresponding to $(1s2l)$ cascade resonances (see Table \ref{table0}).
}
\label{fig_4b}
\end{figure}

In \Fig{fig_4b}, we present the normalized differential cross section \Eq{Eq_diff_norm} for the photon energies corresponding to the four cascade resonances under consideration. The dependence on $\varphi$ can be factorized and parameterized quite accurately as follows:
\begin{eqnarray}
    \frac{d^3 \tilde{\sigma}^{XY}}{d\omega_1 d\Omega_1 d\Omega_2}
    &\approx&\label{Approx1}
    \frac{1}{2\pi}
    \left(C(\omega_1)-A(\omega_1)\sin^2\varphi \right)
    \,.
\end{eqnarray}
The normalization condition yields the following relation between $A$ and $C$: $\pi(2C-A)=1$. 
The values of $C$ and $A$ are listed in Table \ref{table1}.
The unpolarized cross section $\frac{d^3\tilde{\sigma}^{XY}}{d\omega_1 d\Omega_1 d\Omega_2}$ as a function of $\varphi$ has a symmetrical shape relative to the center of [0, $2\pi$]. We note that, for fixed photon and electron polarizations, Eq.(\ref{sks9309d}) does not take place and this symmetry is absent.

\begin{figure}[ht]
\includegraphics[width=20pc]{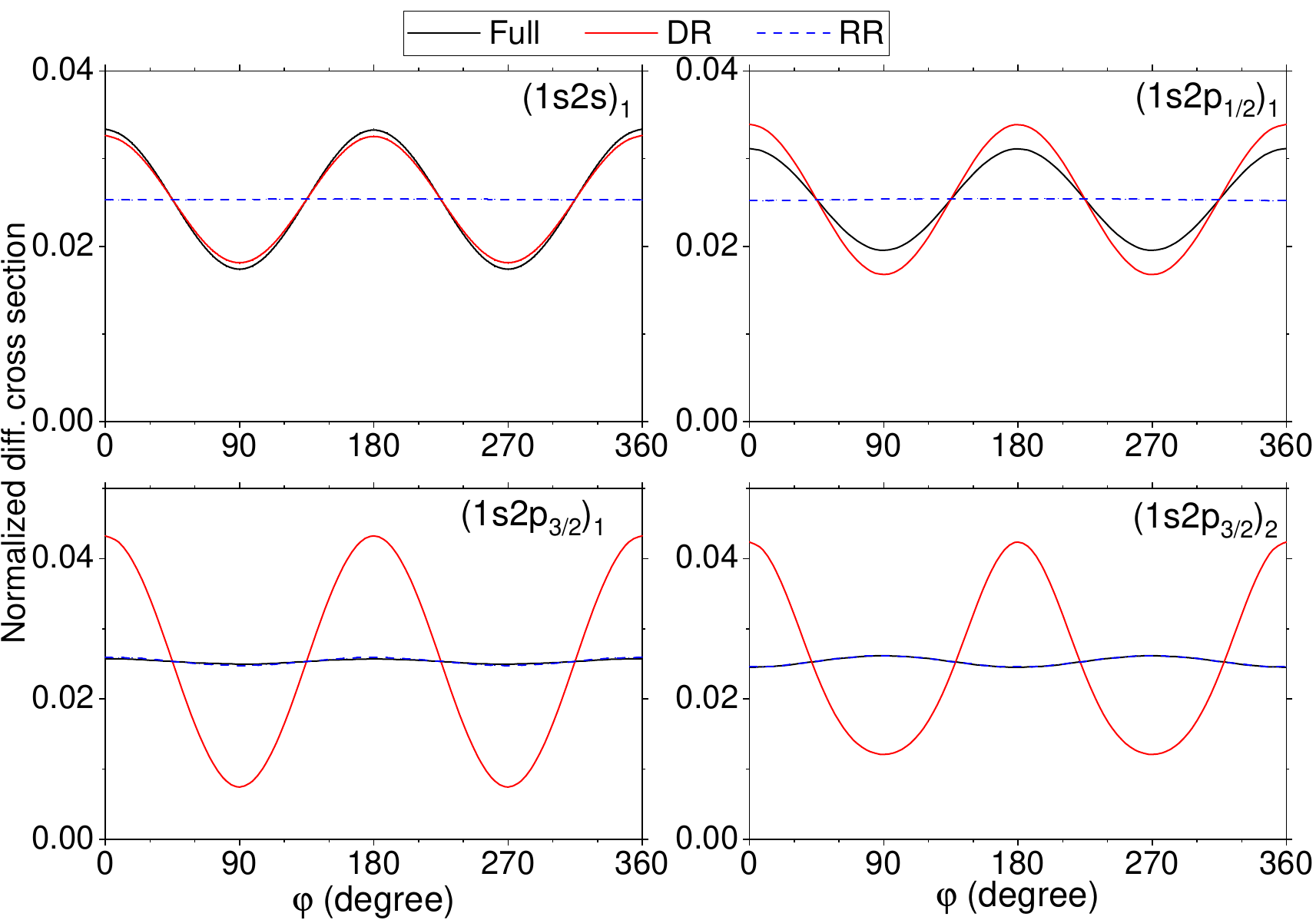}
\caption{(XY-plane) Normalized differential cross section of the two-photon electron capture by H-like uranium ion as 
a function of $\varphi=|\varphi_2-\varphi_1|$ for photon energies corresponding to $(1s2l)$ cascade resonances (see Table \ref{table0}). Black line corresponds to full calculation. Blue and red lines correspond to calculations that take into account only RR and DR, respectively. }
\label{fig_5}
\end{figure}

From Fig. \ref{fig_4b}, we observe that the differential cross sections for the $(1s 2s)_1$ and $(1s2p_{1/2})_1$ cascades exhibit a more pronounced angular dependence compared to those for the $(1s2p_{3/2})_{1,2}$ cascades. This behavior is linked to the role of the DR channel. As discussed in Section \ref{subsec_one_ph_ang_distr}, the DR channel plays a significant role in the $(1s 2s)_1$ and $(1s2p_{1/2})_1$ cascades, while it is strongly suppressed by the RR channel in the $(1s2p_{3/2})_{1,2}$ cascades. To clarify this phenomenon, we present the normalized contributions of the DR and RR channels separately, along with the full calculation, in Fig. \ref{fig_5}.
From Fig. \ref{fig_5}, we can see that the DR exhibits a noticeable angular dependence, whereas the angular dependence for the RR is relatively weak.
Therefore, we conclude that the significant angular dependence observed for the $(1s 2s)_1$ and $(1s2p_{1/2})_1$ cascades is evidence of the DR channel's influence, while the weak angular dependence for the $(1s2p_{3/2})_{1,2}$ cascades indicates the limited role of the DR channel in these cascades.

\begin{table}
\caption{Parameters of the approximation (\ref{Approx1}) for the differential cross section with fixed polar angles of the photon momenta ($\theta_1=\theta_2=\pi/2$).}
\begin{ruledtabular}
\begin{tabular}{ c| c| c| c}
 state & $\frac{d^3\sigma^{XZ}}{d\omega_1 d\cos\theta_1 d\cos\theta_2}$ & C & A \\ 
  & (barn/keV) & ($10^{-2}$ rad$^{-2}$) & ($10^{-2}$ rad$^{-2}$)\\ 
  \hline
 $(1s2s)_1$ & 7.04 $\times 10^5$ & 3.34 & 1.61\\  
 $(1s2p_{1/2})_1$ & 2.28 $\times 10^3$ & 3.11 & 1.16\\ 
 $(1s2p_{3/2})_1$ & 35.4 & 2.57 & 0.076\\  
 $(1s2p_{3/2})_2$ & 4.15 $\times 10^3$ & 2.45 & -0.16
\end{tabular}
\end{ruledtabular}
\label{table1}
\end{table}

\subsubsection{XZ-plane}
Here, we consider photons with momenta in the XZ-plane, so the photon momenta ($\zhk_1$ and $\zhk_2$) and the incident electron momentum ($\zhp$) lie in the same plane.
Accordingly, in this scenario, the azimuthal angles of the emitted photons ($\varphi_i$, $i=1,2$) can take two values: $0^\circ$ and $180^\circ$. The polar angles ($\theta_i$, $i=1,2$) vary within the interval $[0,\pi]$.

It is convenient to introduce an angles $\tilde{\theta}_i$ varying within the interval $[0,2\pi]$
\begin{eqnarray}
    \tilde{\theta}_i
    &=&\label{l04k2}
    \left\{
    \begin{array}{cl}
    \theta_i &, \quad\mbox{for}\quad\varphi_i=0^\circ \\
    2\pi-\theta_i &, \quad\mbox{for}\quad\varphi_i=180^\circ  \\
    \end{array}
    \right.
    \,.
\end{eqnarray}   
We investigate the differential cross section as a function of $\omega_1$, $\tilde{\theta}_1$ and $\tilde{\theta}_2$
\begin{eqnarray}
    &&\nonumber
    \frac{d^3\sigma^{XZ}}{d\omega_1 d\Omega_1 d\Omega_2}(\omega_1, \tilde{\theta}_1, \tilde{\theta}_2)
    \\
    &&
    =\frac{d^3\sigma}{d\omega_1 d\Omega_1 d\Omega_2}\left( \omega_1,\theta_1,\theta_2,|\varphi_2 - \varphi_1| \right)
    \,,
\end{eqnarray}   
where $\tilde{\theta}_i$ is related to $\theta_i$ and $\varphi_i$ in accordance with equation (\ref{l04k2}). 

\begin{figure}
\includegraphics[width=20pc]{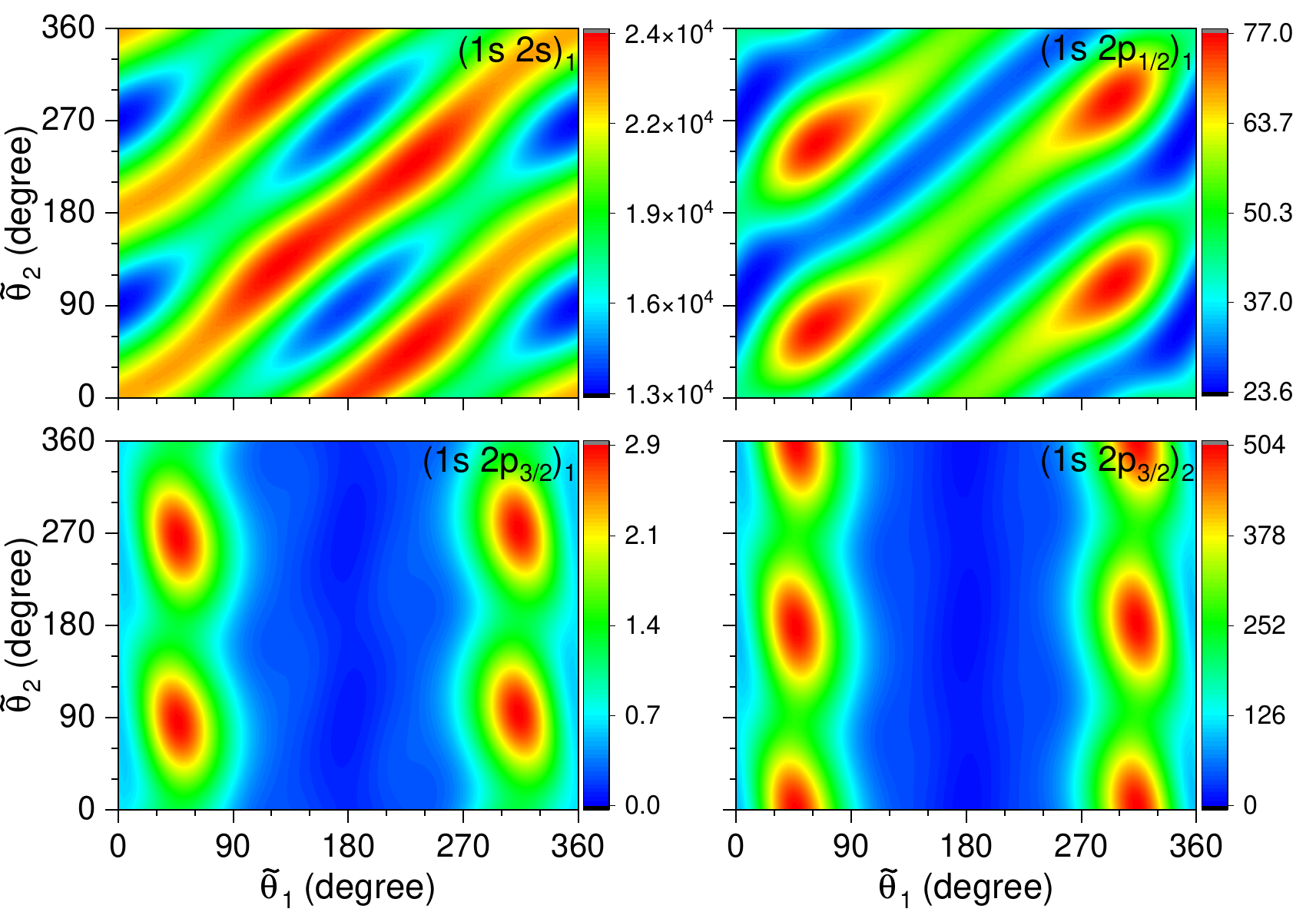}
\caption{(XZ-plane) Differential cross section of the two-photon electron capture by H-like uranium ion as 
a function of $\tilde{\theta}_1$ and $\tilde{\theta}_2$. The photon energies correspond to $(1s2l)$ cascade resonances (see Table \ref{table0}).}
\label{fig_10}
\end{figure}

In Fig. \ref{fig_10}, we present the angular differential cross section for the emitted photons corresponding to the four cascade resonances under consideration. The energies $\omega_1^{(\txt{res},N)}$ associated with these resonances are shown in the Table \ref{table0}.

Due to the azimuthal symmetry of the process relative to the direction of the incident electron momentum, we have the following symmetry in the XZ-plane
\begin{eqnarray}
    &&\nonumber
    \frac{d^3\sigma^{XZ}}{d\omega_1 d\Omega_1 d\Omega_2}(\omega_1, \tilde{\theta}_1, \tilde{\theta}_2)
    \\
    &&
    =\frac{d^3\sigma^{XZ}}{d\omega_1 d\Omega_1 d\Omega_2}(\omega_1, 2\pi-\tilde{\theta}_1, 2\pi-\tilde{\theta}_2)
    \,.
\end{eqnarray}   

The complex structure of the two-photon angular distribution is reflected in the one-photon angular distribution, as discussed in Section~\ref{subsec_one_ph_ang_distr}. For instance, the two-photon distribution for the photon energy corresponding to the $(1s2p_{3/2})_2$ cascade shows two maxima at $\tilde{\theta}_1= 45^\circ$ and $315^\circ$ (for fixed $\tilde{\theta}_2$), and three maxima at $\tilde{\theta}_2 = 0^\circ$, $180^\circ$, and $360^\circ$ (for fixed $\tilde{\theta}_1$). This pattern matches the one-photon distribution shown in Fig. \ref{fig_one_ph_8}, where $\theta_1$ has a maximum at $\theta_1=45^{\circ}$ and $\theta_2$ has two maxima at $\theta_2=0^{\circ}, 180^{\circ}$.

We observe that in cascades where the DR channel dominates ($(1s2s)_1$ and $(1s2p_{1/2})_1$), the emitted photons are significantly more correlated compared to cascades where the RR channel dominates ($(1s2p_{3/2})_1$ and $(1s2p_{3/2})_2$). For instance, in $(1s2s)_1$ cascade, for any given angle $\tilde{\theta}_1$, there is a corresponding $\tilde{\theta}_2$  where the differential cross section shows a pronounced maximum. In contrast, for $(1s2p_{3/2})_1$ cascade, when $\tilde{\theta}_1=180$, the differential cross section remains small for all values of $\tilde{\theta}_2$.

\begin{figure}
\includegraphics[width=20pc]{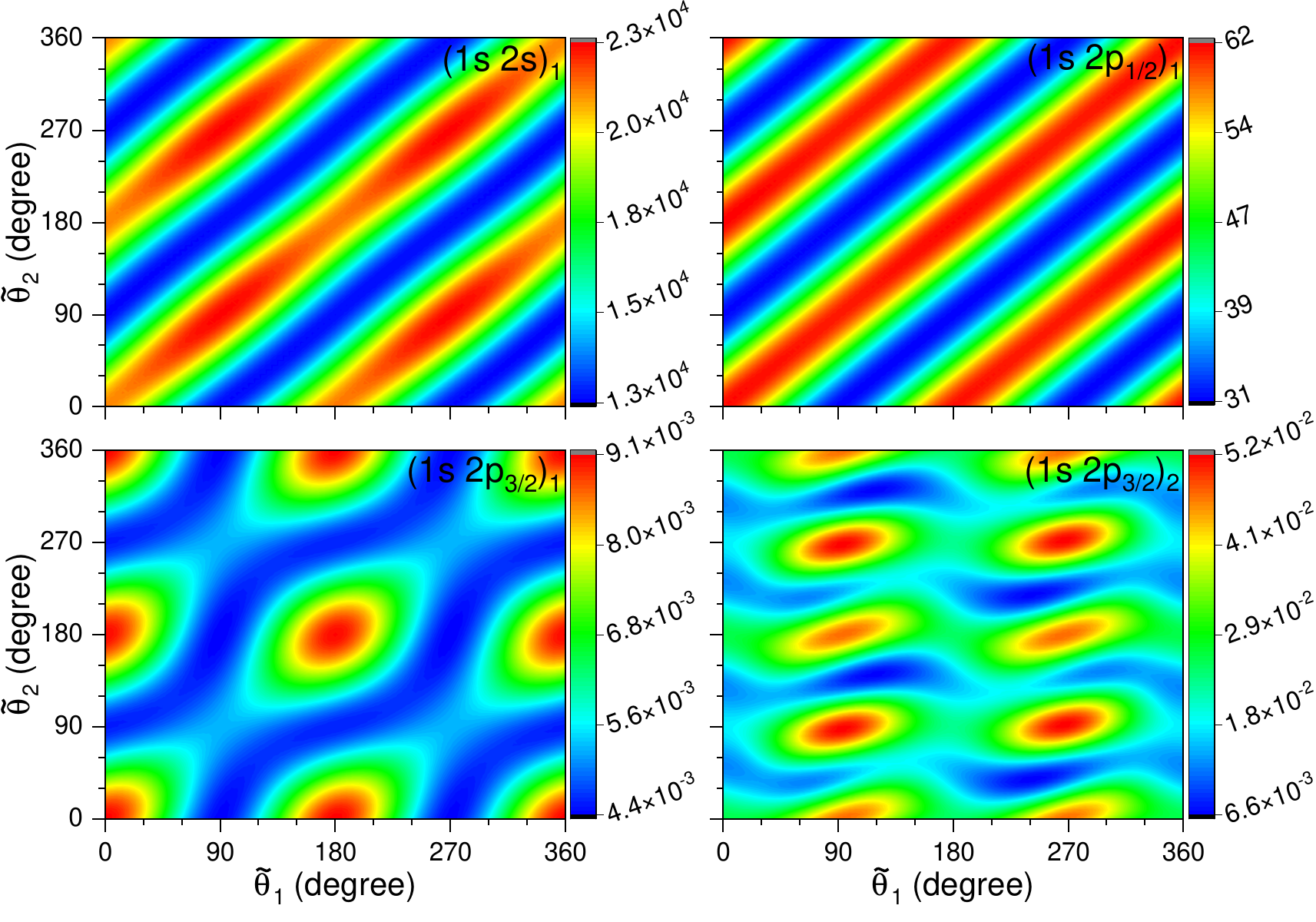}
\caption{(XZ-plane) The same as in Fig. \ref{fig_10}, but only DR channel is taken into account.}
\label{fig_10_DR}
\end{figure}

\begin{figure}
\includegraphics[width=20pc]{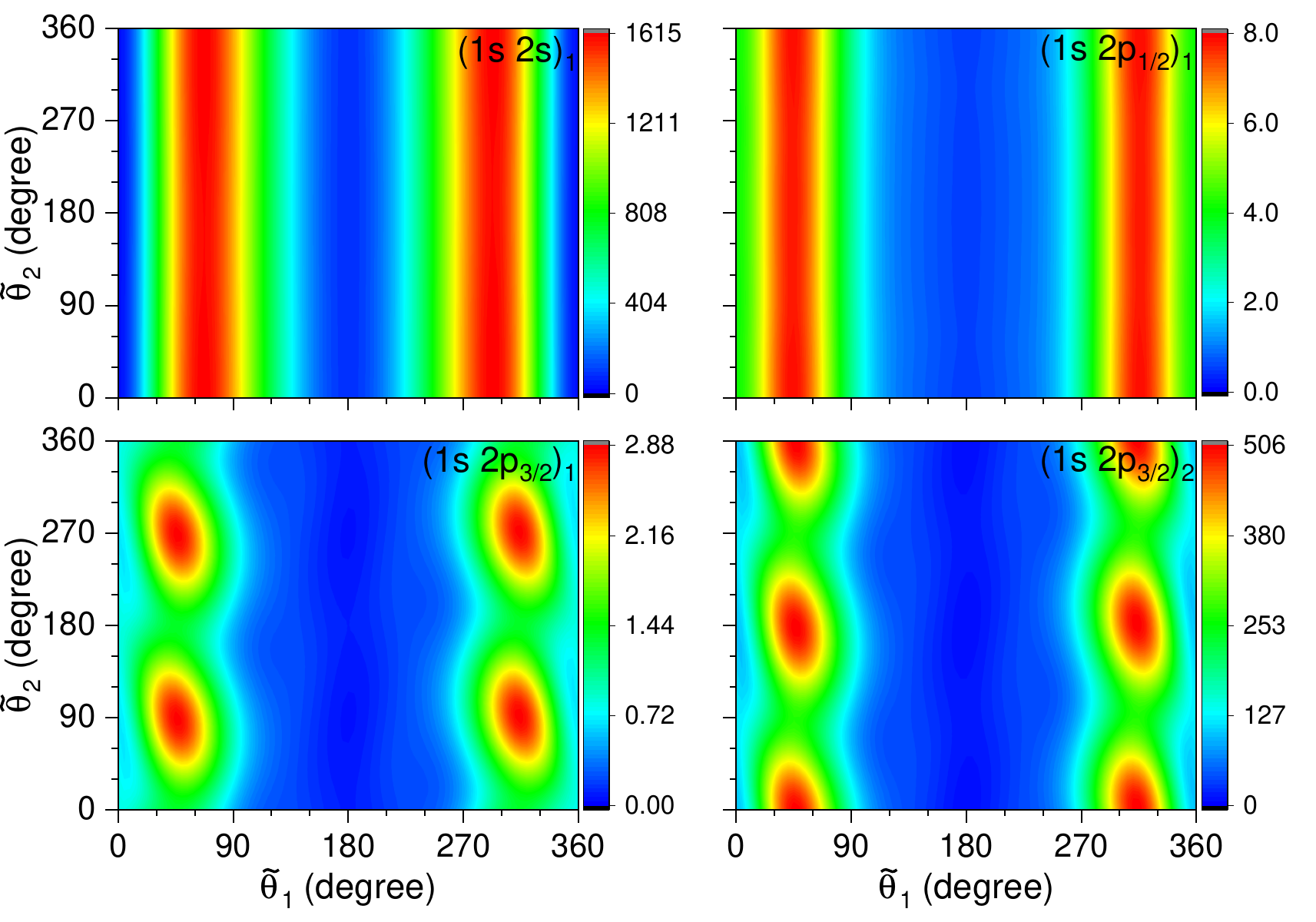}
\caption{(XZ-plane) The same as in Fig. \ref{fig_10}, but only RR channel is taken into account.}
\label{fig_10_RR}
\end{figure}

The obtained angular distribution is shaped by the relative contributions of the DR and RR channels for each cascade resonance. In the upper panels, the DR channel dominates, while in the lower panels, the RR channel contribution is more significant. To illustrate this, we show the individual contributions of the DR and RR channels separately in Figs.~\ref{fig_10_DR} and \ref{fig_10_RR}.

In the DR-dominated cascades, when only the DR channel can be considered (as shown in the upper panels of Fig.\ref{fig_10_DR}), we find that the first photon can be emitted in any direction, but the second photon remains correlated with it. Alternatively, the second photon can be emitted in any direction, and the first photon adjusts to its direction. A similar trend is evident in the RR-dominated cascades (lower panels of Fig.\ref{fig_10_DR}), though to a lesser extent, as the RR resonances are weaker and cannot be fully isolated from DR contributions.

In Fig.~\ref{fig_10_RR}, where only the RR channel contribution is considered, we see a weak dependence on the second photon, reflecting the weak correlation between the first photon, corresponding to the transition from the initial state to the singly excited state, and the second photon, which corresponds to the transition from the singly excited state to the ground state.

\section{Summary}
We investigated the photons angular distribution in resonant two-photon electron capture by H-like uranium. Our study focused on the incident electron energy that corresponds to the largest contribution from the DR channel and the maximum of the total cross section. We conducted a detailed analysis of the main cascade transitions [$(1s2p)$, $(1s2s)$], which contribute most significantly to the cross section. We investigated the one-photon angular distribution, where the integration over the angles of one of the photons was performed. The angular distribution for the resonant photon is in a qualitatively good agreement with results obtained within the single-photon approximation. Going beyond the single-photon approximation, we examined the angular distribution of the satellite photon. 

We analyzed the two-photon angular distribution for two specific geometries: the XY-plane, where the photon momenta are perpendicular to the incident electron momentum, and the XZ-plane, where the incident electron and the photon momenta lie in the same plane. The angular distribution of the photons depends on the relative contributions of the DR and RR channels. The cascade resonances considered fall into two categories: those dominated by the DR channel and those dominated by the RR channel.

In the XY-plane, the DR channel exhibits noticeable oscillations in the angular distribution, while the RR channel leads to a more isotropic pattern, shaping the overall angular dependence in this plane.

In the XZ-plane, for the RR-dominated resonances, the photons associated with the free-bound transition are not aligned with the incident electron momentum, and the second photons show only a weak correlation with the direction of the first photons. On the contrary, for resonances dominated by the DR, the emitted photons correlate with each other, but not with the incident electron momentum.

The presented investigation of the RR and DR channels demonstrates a strong interference between these channels in the two-photon angular distribution.

\newpage
\begin{acknowledgments}
The work of D.Y. is supported by the National Key Research and Development Program of China under Grant No. 2022YFA1602501 and the National Natural Science Foundation of China under Grant No. 12011530060.
The work of K.N.L., O.Y.A. was supported by the Chinese Academy of Sciences (CAS) Presidents International Fellowship Initiative (PIFI) under Grant Nos. 2022VMC0002 and 2025PVA0072, respectively. 
\end{acknowledgments}


\begin{thebibliography}{18}%
\makeatletter
\providecommand \@ifxundefined [1]{%
 \@ifx{#1\undefined}
}%
\providecommand \@ifnum [1]{%
 \ifnum #1\expandafter \@firstoftwo
 \else \expandafter \@secondoftwo
 \fi
}%
\providecommand \@ifx [1]{%
 \ifx #1\expandafter \@firstoftwo
 \else \expandafter \@secondoftwo
 \fi
}%
\providecommand \natexlab [1]{#1}%
\providecommand \enquote  [1]{``#1''}%
\providecommand \bibnamefont  [1]{#1}%
\providecommand \bibfnamefont [1]{#1}%
\providecommand \citenamefont [1]{#1}%
\providecommand \href@noop [0]{\@secondoftwo}%
\providecommand \href [0]{\begingroup \@sanitize@url \@href}%
\providecommand \@href[1]{\@@startlink{#1}\@@href}%
\providecommand \@@href[1]{\endgroup#1\@@endlink}%
\providecommand \@sanitize@url [0]{\catcode `\\12\catcode `\$12\catcode
  `\&12\catcode `\#12\catcode `\^12\catcode `\_12\catcode `\%12\relax}%
\providecommand \@@startlink[1]{}%
\providecommand \@@endlink[0]{}%
\providecommand \url  [0]{\begingroup\@sanitize@url \@url }%
\providecommand \@url [1]{\endgroup\@href {#1}{\urlprefix }}%
\providecommand \urlprefix  [0]{URL }%
\providecommand \Eprint [0]{\href }%
\providecommand \doibase [0]{https://doi.org/}%
\providecommand \selectlanguage [0]{\@gobble}%
\providecommand \bibinfo  [0]{\@secondoftwo}%
\providecommand \bibfield  [0]{\@secondoftwo}%
\providecommand \translation [1]{[#1]}%
\providecommand \BibitemOpen [0]{}%
\providecommand \bibitemStop [0]{}%
\providecommand \bibitemNoStop [0]{.\EOS\space}%
\providecommand \EOS [0]{\spacefactor3000\relax}%
\providecommand \BibitemShut  [1]{\csname bibitem#1\endcsname}%
\let\auto@bib@innerbib\@empty
\bibitem [{\citenamefont {Nakamura}\ \emph {et~al.}(2007)\citenamefont
  {Nakamura}, \citenamefont {Tobiyama}, \citenamefont {Nohara}, \citenamefont
  {Kavanagh}, \citenamefont {Watanabe}, \citenamefont {Sakaue}, \citenamefont
  {Li}, \citenamefont {Kato}, \citenamefont {Currell}, \citenamefont {Yamada},\
  and\ \citenamefont {Ohtani}}]{Nakamura_2007}%
  \BibitemOpen
  \bibfield  {author} {\bibinfo {author} {\bibfnamefont {N.}~\bibnamefont
  {Nakamura}}, \bibinfo {author} {\bibfnamefont {H.}~\bibnamefont {Tobiyama}},
  \bibinfo {author} {\bibfnamefont {H.}~\bibnamefont {Nohara}}, \bibinfo
  {author} {\bibfnamefont {A.~P.}\ \bibnamefont {Kavanagh}}, \bibinfo {author}
  {\bibfnamefont {H.}~\bibnamefont {Watanabe}}, \bibinfo {author}
  {\bibfnamefont {H.~A.}\ \bibnamefont {Sakaue}}, \bibinfo {author}
  {\bibfnamefont {Y.}~\bibnamefont {Li}}, \bibinfo {author} {\bibfnamefont
  {D.}~\bibnamefont {Kato}}, \bibinfo {author} {\bibfnamefont {F.~J.}\
  \bibnamefont {Currell}}, \bibinfo {author} {\bibfnamefont {C.}~\bibnamefont
  {Yamada}},\ and\ \bibinfo {author} {\bibfnamefont {S.}~\bibnamefont
  {Ohtani}},\ }\bibfield  {title} {\bibinfo {title} {Resonant electron
  processes with open-shell highly charged ion targets},\ }\href
  {https://doi.org/10.1088/1742-6596/58/1/058} {\bibfield  {journal} {\bibinfo
  {journal} {Journal of Physics: Conference Series}\ }\textbf {\bibinfo
  {volume} {58}},\ \bibinfo {pages} {267} (\bibinfo {year} {2007})}\BibitemShut
  {NoStop}%
\bibitem [{\citenamefont {Yang}\ \emph {et~al.}(2021)\citenamefont {Yang},
  \citenamefont {Gao}, \citenamefont {Yan}, \citenamefont {Yao}, \citenamefont
  {Yang}, \citenamefont {Wu},\ and\ \citenamefont {Hu}}]{Yang2021pra}%
  \BibitemOpen
  \bibfield  {author} {\bibinfo {author} {\bibfnamefont {Z.}~\bibnamefont
  {Yang}}, \bibinfo {author} {\bibfnamefont {J.}~\bibnamefont {Gao}}, \bibinfo
  {author} {\bibfnamefont {W.}~\bibnamefont {Yan}}, \bibinfo {author}
  {\bibfnamefont {K.}~\bibnamefont {Yao}}, \bibinfo {author} {\bibfnamefont
  {J.}~\bibnamefont {Yang}}, \bibinfo {author} {\bibfnamefont {Z.}~\bibnamefont
  {Wu}},\ and\ \bibinfo {author} {\bibfnamefont {Z.}~\bibnamefont {Hu}},\
  }\bibfield  {title} {\bibinfo {title} {Anisotropy and polarization of x-ray
  line emissions in the dielectronic recombination of hydrogenlike
  ${\mathrm{fe}}^{25+}$ ions},\ }\href
  {https://doi.org/10.1103/PhysRevA.104.022809} {\bibfield  {journal} {\bibinfo
   {journal} {Phys. Rev. A}\ }\textbf {\bibinfo {volume} {104}},\ \bibinfo
  {pages} {022809} (\bibinfo {year} {2021})}\BibitemShut {NoStop}%
\bibitem [{\citenamefont {Nakamura}\ \emph {et~al.}(2023)\citenamefont
  {Nakamura}, \citenamefont {Numadate}, \citenamefont {Oishi}, \citenamefont
  {Tong}, \citenamefont {Gao}, \citenamefont {Kato}, \citenamefont {Odaka},
  \citenamefont {Takahashi}, \citenamefont {Tsuzuki}, \citenamefont {Uchida},
  \citenamefont {Watanabe}, \citenamefont {Watanabe},\ and\ \citenamefont
  {Yoneda}}]{Nakamura2023PhysRevLett.130.113001}%
  \BibitemOpen
  \bibfield  {author} {\bibinfo {author} {\bibfnamefont {N.}~\bibnamefont
  {Nakamura}}, \bibinfo {author} {\bibfnamefont {N.}~\bibnamefont {Numadate}},
  \bibinfo {author} {\bibfnamefont {S.}~\bibnamefont {Oishi}}, \bibinfo
  {author} {\bibfnamefont {X.-M.}\ \bibnamefont {Tong}}, \bibinfo {author}
  {\bibfnamefont {X.}~\bibnamefont {Gao}}, \bibinfo {author} {\bibfnamefont
  {D.}~\bibnamefont {Kato}}, \bibinfo {author} {\bibfnamefont {H.}~\bibnamefont
  {Odaka}}, \bibinfo {author} {\bibfnamefont {T.}~\bibnamefont {Takahashi}},
  \bibinfo {author} {\bibfnamefont {Y.}~\bibnamefont {Tsuzuki}}, \bibinfo
  {author} {\bibfnamefont {Y.}~\bibnamefont {Uchida}}, \bibinfo {author}
  {\bibfnamefont {H.}~\bibnamefont {Watanabe}}, \bibinfo {author}
  {\bibfnamefont {S.}~\bibnamefont {Watanabe}},\ and\ \bibinfo {author}
  {\bibfnamefont {H.}~\bibnamefont {Yoneda}},\ }\bibfield  {title} {\bibinfo
  {title} {{Strong Polarization of a $J=1/2$ to $1/2$ Transition Arising from
  Unexpectedly Large Quantum Interference}},\ }\href
  {https://doi.org/10.1103/PhysRevLett.130.113001} {\bibfield  {journal}
  {\bibinfo  {journal} {Phys. Rev. Lett.}\ }\textbf {\bibinfo {volume} {130}},\
  \bibinfo {pages} {113001} (\bibinfo {year} {2023})}\BibitemShut {NoStop}%
\bibitem [{\citenamefont {Lyashchenko}\ \emph {et~al.}(2024)\citenamefont
  {Lyashchenko}, \citenamefont {Andreev},\ and\ \citenamefont
  {Yu}}]{lyashchenko2024}%
  \BibitemOpen
  \bibfield  {author} {\bibinfo {author} {\bibfnamefont {K.~N.}\ \bibnamefont
  {Lyashchenko}}, \bibinfo {author} {\bibfnamefont {O.~Y.}\ \bibnamefont
  {Andreev}},\ and\ \bibinfo {author} {\bibfnamefont {D.}~\bibnamefont {Yu}},\
  }\bibfield  {title} {\bibinfo {title} {Two-photon electron capture by h-like
  uranium},\ }\href {https://doi.org/10.1103/PhysRevA.109.032805} {\bibfield
  {journal} {\bibinfo  {journal} {Phys. Rev. A}\ }\textbf {\bibinfo {volume}
  {109}},\ \bibinfo {pages} {032805} (\bibinfo {year} {2024})}\BibitemShut
  {NoStop}%
\bibitem [{\citenamefont {Eichler}\ and\ \citenamefont
  {St\"{o}hlker}(2007)}]{eichler07}%
  \BibitemOpen
  \bibfield  {author} {\bibinfo {author} {\bibfnamefont {J.}~\bibnamefont
  {Eichler}}\ and\ \bibinfo {author} {\bibfnamefont {T.}~\bibnamefont
  {St\"{o}hlker}},\ }\bibfield  {title} {\bibinfo {title} {{Radiative electron
  capture in relativistic ion-atom collisions and the photoelectric effect in
  hydrogen-like high-Z systems}},\ }\href
  {https://doi.org/https://doi.org/10.1016/j.physrep.2006.11.003} {\bibfield
  {journal} {\bibinfo  {journal} {Phys. Rep.}\ }\textbf {\bibinfo {volume}
  {439}},\ \bibinfo {pages} {1} (\bibinfo {year} {2007})}\BibitemShut {NoStop}%
\bibitem [{\citenamefont {Eichler}\ \emph {et~al.}(1995)\citenamefont
  {Eichler}, \citenamefont {Ichihara},\ and\ \citenamefont
  {Shirai}}]{eichler1995PhysRevA.51.3027}%
  \BibitemOpen
  \bibfield  {author} {\bibinfo {author} {\bibfnamefont {J.}~\bibnamefont
  {Eichler}}, \bibinfo {author} {\bibfnamefont {A.}~\bibnamefont {Ichihara}},\
  and\ \bibinfo {author} {\bibfnamefont {T.}~\bibnamefont {Shirai}},\
  }\bibfield  {title} {\bibinfo {title} {Photon angular distributions from
  radiative electron capture in relativistic atomic collisions},\ }\href
  {https://doi.org/10.1103/PhysRevA.51.3027} {\bibfield  {journal} {\bibinfo
  {journal} {Phys. Rev. A}\ }\textbf {\bibinfo {volume} {51}},\ \bibinfo
  {pages} {3027} (\bibinfo {year} {1995})}\BibitemShut {NoStop}%
\bibitem [{\citenamefont {Karasiov}\ \emph {et~al.}(1992)\citenamefont
  {Karasiov}, \citenamefont {Labzowsky}, \citenamefont {Nefiodov},\ and\
  \citenamefont {Shabaev}}]{karasiov92p453}%
  \BibitemOpen
  \bibfield  {author} {\bibinfo {author} {\bibfnamefont {V.}~\bibnamefont
  {Karasiov}}, \bibinfo {author} {\bibfnamefont {L.}~\bibnamefont {Labzowsky}},
  \bibinfo {author} {\bibfnamefont {A.}~\bibnamefont {Nefiodov}},\ and\
  \bibinfo {author} {\bibfnamefont {V.}~\bibnamefont {Shabaev}},\ }\bibfield
  {title} {\bibinfo {title} {Overlapping resonances in the process of
  recombination of an electron with hydrogenlike uranium},\ }\href
  {https://doi.org/https://doi.org/10.1016/0375-9601(92)90687-H} {\bibfield
  {journal} {\bibinfo  {journal} {Physics Letters A}\ }\textbf {\bibinfo
  {volume} {161}},\ \bibinfo {pages} {453} (\bibinfo {year}
  {1992})}\BibitemShut {NoStop}%
\bibitem [{\citenamefont {Bernhardt}\ \emph {et~al.}(2011)\citenamefont
  {Bernhardt}, \citenamefont {Brandau}, \citenamefont {Harman}, \citenamefont
  {Kozhuharov}, \citenamefont {M\"uller}, \citenamefont {Scheid}, \citenamefont
  {Schippers}, \citenamefont {Schmidt}, \citenamefont {Yu}, \citenamefont
  {Artemyev}, \citenamefont {Tupitsyn}, \citenamefont {B\"ohm}, \citenamefont
  {Bosch}, \citenamefont {Currell}, \citenamefont {Franzke}, \citenamefont
  {Gumberidze}, \citenamefont {Jacobi}, \citenamefont {Mokler}, \citenamefont
  {Nolden}, \citenamefont {Spillman}, \citenamefont {Stachura}, \citenamefont
  {Steck},\ and\ \citenamefont {St\"ohlker}}]{bernhardt11}%
  \BibitemOpen
  \bibfield  {author} {\bibinfo {author} {\bibfnamefont {D.}~\bibnamefont
  {Bernhardt}}, \bibinfo {author} {\bibfnamefont {C.}~\bibnamefont {Brandau}},
  \bibinfo {author} {\bibfnamefont {Z.}~\bibnamefont {Harman}}, \bibinfo
  {author} {\bibfnamefont {C.}~\bibnamefont {Kozhuharov}}, \bibinfo {author}
  {\bibfnamefont {A.}~\bibnamefont {M\"uller}}, \bibinfo {author}
  {\bibfnamefont {W.}~\bibnamefont {Scheid}}, \bibinfo {author} {\bibfnamefont
  {S.}~\bibnamefont {Schippers}}, \bibinfo {author} {\bibfnamefont {E.~W.}\
  \bibnamefont {Schmidt}}, \bibinfo {author} {\bibfnamefont {D.}~\bibnamefont
  {Yu}}, \bibinfo {author} {\bibfnamefont {A.~N.}\ \bibnamefont {Artemyev}},
  \bibinfo {author} {\bibfnamefont {I.~I.}\ \bibnamefont {Tupitsyn}}, \bibinfo
  {author} {\bibfnamefont {S.}~\bibnamefont {B\"ohm}}, \bibinfo {author}
  {\bibfnamefont {F.}~\bibnamefont {Bosch}}, \bibinfo {author} {\bibfnamefont
  {F.~J.}\ \bibnamefont {Currell}}, \bibinfo {author} {\bibfnamefont
  {B.}~\bibnamefont {Franzke}}, \bibinfo {author} {\bibfnamefont
  {A.}~\bibnamefont {Gumberidze}}, \bibinfo {author} {\bibfnamefont
  {J.}~\bibnamefont {Jacobi}}, \bibinfo {author} {\bibfnamefont {P.~H.}\
  \bibnamefont {Mokler}}, \bibinfo {author} {\bibfnamefont {F.}~\bibnamefont
  {Nolden}}, \bibinfo {author} {\bibfnamefont {U.}~\bibnamefont {Spillman}},
  \bibinfo {author} {\bibfnamefont {Z.}~\bibnamefont {Stachura}}, \bibinfo
  {author} {\bibfnamefont {M.}~\bibnamefont {Steck}},\ and\ \bibinfo {author}
  {\bibfnamefont {T.}~\bibnamefont {St\"ohlker}},\ }\bibfield  {title}
  {\bibinfo {title} {Breit interaction in dielectronic recombination of
  hydrogenlike uranium},\ }\href {https://doi.org/10.1103/PhysRevA.83.020701}
  {\bibfield  {journal} {\bibinfo  {journal} {Phys. Rev. A}\ }\textbf {\bibinfo
  {volume} {83}},\ \bibinfo {pages} {020701} (\bibinfo {year}
  {2011})}\BibitemShut {NoStop}%
\bibitem [{\citenamefont {Lyashchenko}\ and\ \citenamefont
  {Andreev}(2015)}]{lyashchenko2015}%
  \BibitemOpen
  \bibfield  {author} {\bibinfo {author} {\bibfnamefont {K.~N.}\ \bibnamefont
  {Lyashchenko}}\ and\ \bibinfo {author} {\bibfnamefont {O.~Y.}\ \bibnamefont
  {Andreev}},\ }\bibfield  {title} {\bibinfo {title} {{Importance of the Breit
  interaction for calculation of the differential cross section for
  dielectronic recombination with one-electron uranium}},\ }\href
  {https://doi.org/10.1103/PhysRevA.91.012511} {\bibfield  {journal} {\bibinfo
  {journal} {Phys. Rev. A}\ }\textbf {\bibinfo {volume} {91}},\ \bibinfo
  {pages} {012511} (\bibinfo {year} {2015})}\BibitemShut {NoStop}%
\bibitem [{\citenamefont {Zakowicz}\ \emph {et~al.}(2004)\citenamefont
  {Zakowicz}, \citenamefont {Scheid},\ and\ \citenamefont
  {Grün}}]{zakowicz04}%
  \BibitemOpen
  \bibfield  {author} {\bibinfo {author} {\bibfnamefont {S.}~\bibnamefont
  {Zakowicz}}, \bibinfo {author} {\bibfnamefont {W.}~\bibnamefont {Scheid}},\
  and\ \bibinfo {author} {\bibfnamefont {N.}~\bibnamefont {Grün}},\ }\bibfield
   {title} {\bibinfo {title} {Dielectronic recombination into hydrogen-like
  heavy ions with emission of two photons},\ }\href
  {https://doi.org/10.1088/0953-4075/37/1/008} {\bibfield  {journal} {\bibinfo
  {journal} {Journal of Physics B: Atomic, Molecular and Optical Physics}\
  }\textbf {\bibinfo {volume} {37}},\ \bibinfo {pages} {131} (\bibinfo {year}
  {2004})}\BibitemShut {NoStop}%
\bibitem [{\citenamefont {Nakamura}\ \emph {et~al.}(2008)\citenamefont
  {Nakamura}, \citenamefont {Kavanagh}, \citenamefont {Watanabe}, \citenamefont
  {Sakaue}, \citenamefont {Li}, \citenamefont {Kato}, \citenamefont {Currell},\
  and\ \citenamefont {Ohtani}}]{Nakamura2008PhysRevLett.100.073203}%
  \BibitemOpen
  \bibfield  {author} {\bibinfo {author} {\bibfnamefont {N.}~\bibnamefont
  {Nakamura}}, \bibinfo {author} {\bibfnamefont {A.~P.}\ \bibnamefont
  {Kavanagh}}, \bibinfo {author} {\bibfnamefont {H.}~\bibnamefont {Watanabe}},
  \bibinfo {author} {\bibfnamefont {H.~A.}\ \bibnamefont {Sakaue}}, \bibinfo
  {author} {\bibfnamefont {Y.}~\bibnamefont {Li}}, \bibinfo {author}
  {\bibfnamefont {D.}~\bibnamefont {Kato}}, \bibinfo {author} {\bibfnamefont
  {F.~J.}\ \bibnamefont {Currell}},\ and\ \bibinfo {author} {\bibfnamefont
  {S.}~\bibnamefont {Ohtani}},\ }\bibfield  {title} {\bibinfo {title} {Evidence
  for strong breit interaction in dielectronic recombination of highly charged
  heavy ions},\ }\href {https://doi.org/10.1103/PhysRevLett.100.073203}
  {\bibfield  {journal} {\bibinfo  {journal} {Phys. Rev. Lett.}\ }\textbf
  {\bibinfo {volume} {100}},\ \bibinfo {pages} {073203} (\bibinfo {year}
  {2008})}\BibitemShut {NoStop}%
\bibitem [{\citenamefont {Mahmood}\ \emph {et~al.}(2012)\citenamefont
  {Mahmood}, \citenamefont {Ali}, \citenamefont {Orban}, \citenamefont
  {Tashenov}, \citenamefont {Lindroth},\ and\ \citenamefont
  {Schuch}}]{Mahmood_2012}%
  \BibitemOpen
  \bibfield  {author} {\bibinfo {author} {\bibfnamefont {S.}~\bibnamefont
  {Mahmood}}, \bibinfo {author} {\bibfnamefont {S.}~\bibnamefont {Ali}},
  \bibinfo {author} {\bibfnamefont {I.}~\bibnamefont {Orban}}, \bibinfo
  {author} {\bibfnamefont {S.}~\bibnamefont {Tashenov}}, \bibinfo {author}
  {\bibfnamefont {E.}~\bibnamefont {Lindroth}},\ and\ \bibinfo {author}
  {\bibfnamefont {R.}~\bibnamefont {Schuch}},\ }\bibfield  {title} {\bibinfo
  {title} {{Recombination and electron impact excitation rate coefficients for
  S XV and S XVI}},\ }\href {https://doi.org/10.1088/0004-637X/754/2/86}
  {\bibfield  {journal} {\bibinfo  {journal} {The Astrophysical Journal}\
  }\textbf {\bibinfo {volume} {754}},\ \bibinfo {pages} {86} (\bibinfo {year}
  {2012})}\BibitemShut {NoStop}%
\bibitem [{\citenamefont {Lindroth}\ \emph {et~al.}(2020)\citenamefont
  {Lindroth}, \citenamefont {Orban}, \citenamefont {Trotsenko},\ and\
  \citenamefont {Schuch}}]{Lindroth2020PhysRevA.101.062706}%
  \BibitemOpen
  \bibfield  {author} {\bibinfo {author} {\bibfnamefont {E.}~\bibnamefont
  {Lindroth}}, \bibinfo {author} {\bibfnamefont {I.}~\bibnamefont {Orban}},
  \bibinfo {author} {\bibfnamefont {S.}~\bibnamefont {Trotsenko}},\ and\
  \bibinfo {author} {\bibfnamefont {R.}~\bibnamefont {Schuch}},\ }\bibfield
  {title} {\bibinfo {title} {{Electron-impact recombination and excitation
  rates for charge-state-selected highly charged Si ions}},\ }\href
  {https://doi.org/10.1103/PhysRevA.101.062706} {\bibfield  {journal} {\bibinfo
   {journal} {Phys. Rev. A}\ }\textbf {\bibinfo {volume} {101}},\ \bibinfo
  {pages} {062706} (\bibinfo {year} {2020})}\BibitemShut {NoStop}%
\bibitem [{\citenamefont {Hu}\ \emph {et~al.}(2022)\citenamefont {Hu},
  \citenamefont {Xiong}, \citenamefont {He}, \citenamefont {Yang},
  \citenamefont {Numadate}, \citenamefont {Huang}, \citenamefont {Yang},
  \citenamefont {Yao}, \citenamefont {Wei}, \citenamefont {Zou}, \citenamefont
  {Wu}, \citenamefont {Ma}, \citenamefont {Wu}, \citenamefont {Gao},\ and\
  \citenamefont {Nakamura}}]{hu2022PhysRevA.105.L030801}%
  \BibitemOpen
  \bibfield  {author} {\bibinfo {author} {\bibfnamefont {Z.}~\bibnamefont
  {Hu}}, \bibinfo {author} {\bibfnamefont {G.}~\bibnamefont {Xiong}}, \bibinfo
  {author} {\bibfnamefont {Z.}~\bibnamefont {He}}, \bibinfo {author}
  {\bibfnamefont {Z.}~\bibnamefont {Yang}}, \bibinfo {author} {\bibfnamefont
  {N.}~\bibnamefont {Numadate}}, \bibinfo {author} {\bibfnamefont
  {C.}~\bibnamefont {Huang}}, \bibinfo {author} {\bibfnamefont
  {J.}~\bibnamefont {Yang}}, \bibinfo {author} {\bibfnamefont {K.}~\bibnamefont
  {Yao}}, \bibinfo {author} {\bibfnamefont {B.}~\bibnamefont {Wei}}, \bibinfo
  {author} {\bibfnamefont {Y.}~\bibnamefont {Zou}}, \bibinfo {author}
  {\bibfnamefont {C.}~\bibnamefont {Wu}}, \bibinfo {author} {\bibfnamefont
  {Y.}~\bibnamefont {Ma}}, \bibinfo {author} {\bibfnamefont {Y.}~\bibnamefont
  {Wu}}, \bibinfo {author} {\bibfnamefont {X.}~\bibnamefont {Gao}},\ and\
  \bibinfo {author} {\bibfnamefont {N.}~\bibnamefont {Nakamura}},\ }\bibfield
  {title} {\bibinfo {title} {Giant retardation effect in electron-electron
  interaction},\ }\href {https://doi.org/10.1103/PhysRevA.105.L030801}
  {\bibfield  {journal} {\bibinfo  {journal} {Phys. Rev. A}\ }\textbf {\bibinfo
  {volume} {105}},\ \bibinfo {pages} {L030801} (\bibinfo {year}
  {2022})}\BibitemShut {NoStop}%
\bibitem [{\citenamefont {{F. Aharonian et al. (Hitomi
  Collaboration)}}(2018)}]{hitomi2018}%
  \BibitemOpen
  \bibfield  {author} {\bibinfo {author} {\bibnamefont {{F. Aharonian et al.
  (Hitomi Collaboration)}}},\ }\bibfield  {title} {\bibinfo {title} {{Atomic
  data and spectral modeling constraints from high-resolution X-ray
  observations of the Perseus cluster with Hitomi*}},\ }\href
  {https://doi.org/10.1093/pasj/psx156} {\bibfield  {journal} {\bibinfo
  {journal} {Publications of the Astronomical Society of Japan}\ }\textbf
  {\bibinfo {volume} {70}},\ \bibinfo {pages} {12} (\bibinfo {year}
  {2018})}\BibitemShut {NoStop}%
\bibitem [{\citenamefont {Andreev}\ \emph {et~al.}(2008)\citenamefont
  {Andreev}, \citenamefont {Labzowsky}, \citenamefont {Plunien},\ and\
  \citenamefont {Solovyev}}]{andreev08pr}%
  \BibitemOpen
  \bibfield  {author} {\bibinfo {author} {\bibfnamefont {O.~Y.}\ \bibnamefont
  {Andreev}}, \bibinfo {author} {\bibfnamefont {L.~N.}\ \bibnamefont
  {Labzowsky}}, \bibinfo {author} {\bibfnamefont {G.}~\bibnamefont {Plunien}},\
  and\ \bibinfo {author} {\bibfnamefont {D.~A.}\ \bibnamefont {Solovyev}},\
  }\bibfield  {title} {\bibinfo {title} {{QED theory of the spectral line
  profile and its applications to atoms and ions}},\ }\href
  {https://doi.org/https://doi.org/10.1016/j.physrep.2007.10.003} {\bibfield
  {journal} {\bibinfo  {journal} {Physics Reports}\ }\textbf {\bibinfo {volume}
  {455}},\ \bibinfo {pages} {135} (\bibinfo {year} {2008})}\BibitemShut
  {NoStop}%
\bibitem [{\citenamefont {Varshalovich}\ \emph {et~al.}(1988)\citenamefont
  {Varshalovich}, \citenamefont {Moskalev},\ and\ \citenamefont
  {Khersonskii}}]{Varshalovich1988QuantumTO}%
  \BibitemOpen
  \bibfield  {author} {\bibinfo {author} {\bibfnamefont {D.~A.}\ \bibnamefont
  {Varshalovich}}, \bibinfo {author} {\bibfnamefont {A.~N.}\ \bibnamefont
  {Moskalev}},\ and\ \bibinfo {author} {\bibfnamefont {V.~K.}\ \bibnamefont
  {Khersonskii}},\ }\href {https://doi.org/10.1142/0270} {\emph {\bibinfo
  {title} {Quantum Theory of Angular Momentum}}}\ (\bibinfo  {publisher} {World
  Scientific Publishing Co. Pte. Ltd.},\ \bibinfo {address} {Singapore},\
  \bibinfo {year} {1988})\BibitemShut {NoStop}%
\bibitem [{\citenamefont {\mbox{Yu}. Andreev}\ \emph
  {et~al.}(2009)\citenamefont {\mbox{Yu}. Andreev}, \citenamefont {Labzowsky},\
  and\ \citenamefont {Prigorovsky}}]{andreev09p042514}%
  \BibitemOpen
  \bibfield  {author} {\bibinfo {author} {\bibfnamefont {O.}~\bibnamefont
  {\mbox{Yu}. Andreev}}, \bibinfo {author} {\bibfnamefont {L.~N.}\ \bibnamefont
  {Labzowsky}},\ and\ \bibinfo {author} {\bibfnamefont {A.~V.}\ \bibnamefont
  {Prigorovsky}},\ }\bibfield  {title} {\bibinfo {title} {Line-profile approach
  to the description of the electron-recombination process for the highly
  charged ions},\ }\href {https://doi.org/10.1103/PhysRevA.80.042514}
  {\bibfield  {journal} {\bibinfo  {journal} {Phys. Rev. A}\ }\textbf {\bibinfo
  {volume} {80}},\ \bibinfo {pages} {042514} (\bibinfo {year}
  {2009})}\BibitemShut {NoStop}%
\end{thebibliography}

%

\end{document}